\documentclass[%
pre,preprint,superscriptaddress,
tightenlines,
showpacs,
twocolumn,
a4paper,12pt
]{revtex4}
\usepackage{amssymb,amsmath}
\usepackage{graphicx}
\usepackage{subfigure}

\renewcommand{\thesubfigure}{(\alph{subfigure})}

 \makeatletter
 \renewcommand{\@thesubfigure}{\thesubfigure\space}
 \makeatother

\newcommand{\sca}[2]{\ensuremath{\bigl({#1},{#2}\bigr)}}
\newcommand{\avr}[1]{\ensuremath{\langle{#1}\rangle}}

\newcommand{\cnj}[1]{{#1}^{\ast}}
\newcommand{\hcnj}[1]{{#1}^{+}}

\newcommand{\bs}[1]{\boldsymbol{#1}}
\newcommand{\vc}[1]{\mathbf{#1}}

\newcommand{\dd}{\mathrm{d}}
\newcommand{\DD}{\ensuremath{\mathcal{D}}}

\newcommand{\uv}{\mathrm{uv}}
\newcommand{\smx}{\mathrm{max}}
\newcommand{\smn}{\mathrm{min}}

\newcommand{\prt}[1]{\partial_{#1}}

\newcommand{\diag}{\mathop{\rm diag}\nolimits}
\newcommand{\bnbl}{\boldsymbol{\nabla}}

\begin{document}


 \title{\bfseries Saddle-splay term induced orientational instability in 
nematic liquid crystal cells and director fluctuations at substrates 
}

\author{A.D.~Kiselev}
\email[Email address: ]{kisel@elit.chernigov.ua}
\affiliation{%
 Chernigov State Technological University,
 Shevchenko Street 95,
 14027 Chernigov, Ukraine
}

\date{\today}

\begin{abstract}
  We analyze stability of the planar orientational structure in a
  nematic liquid crystal (NLC) cell with planar anchoring conditions
  at both substrates.  Specifically, we study the instabilities of the
  ground state caused by surface elasticity at large saddle-splay
  elastic constant, $K_{24}$.  We show that such instabilities are
  induced by the director fluctuations at confining walls and derive
  the surface part of static correlation functions as a functional
  integral over these fluctuations characterized by the surface free
  energy.  From the surface part of the correlator we derive the
  stability conditions for the planar structure with respect to the
  fluctuation modes characterized by the in-plane wavenumbers and by
  the parity symmetry.  These conditions are analyzed in the cell
  thickness~--~fluctuation wavelength plane by using the parameterization
  for the boundary curve of the instability region.
  For relatively small $K_{24}$ violating the Ericksen inequalities 
$0 < K_{24} < 2 \min(K_{1},K_{2})$ 
  the theory predicts that the critical fluctuation mode of the
  wavelength, $\lambda_c$, will render the structure unstable when the
  thickness of the cell is below its critical value, $d_c$.  The
  parity of the critical mode changes as the twist-splay ratio $K_2/K_1$ is
  passing through the unity. Further increase of $K_{24}$ beyond the
  second threshold value, $4K_1 K_2/(K_1+K_2)$, leads to the
  instability with respect to the short wavelength fluctuations
  regardless of the cell thickness.  We compute the critical thickness
  and the critical wavelength as functions of $K_{24}$, the twist-splay
  ratio and the azimuthal anchoring strength.

\end{abstract}

\pacs{%
61.30.Gd,  64.70.Md, 61.30.Hn
}

\keywords{%
nematic liquid crystal; surface elasticity; periodic director deformations
}

\maketitle

\section{Introduction}
\label{sec:intro}

Nematic liquid crystals (NLC) 
confined in restricted geometries are technologically
important~\cite{Doan:1990} and have been the subject of intense
studies over the past few decades~\cite{Craw:mpl:1993,Craw:bk:1996}.
Anisotropy of the vast majority of NLCs is locally uniaxial and
molecules of a NLC align on average along a local unit director.
Orientational structures in NLCs are thus defined
by distributions of the director $\vc{n}(\vc{r})$ and
the well-established continuum elastic theory 
provides the phenomenological description of orientational 
distortions~\cite{Gennes:bk:1993,Luben:bk:1995}.

In the absence of external fields, orientational structures in 
spatially bounded
NLCs crucially depend on the conditions at confining
walls. These are macroscopically characterized by the surface
contribution to the elastic free energy,
$F_s$, that adds to the Frank elastic energy, $F_b$,
describing elasticity of NLC in the bulk, 
to yield the total elastic free energy of  a NLC
in the presence of the confining surfaces
\begin{align}
\label{eq:Fn-gen}
&
F[\vc{n}]=F_b[\vc{n}]+F_s[\vc{n}],
\end{align}
\begin{align}
\label{eq:fb-gen}
&
  F_b=\frac{1}{2}\, \int_V\Bigl[
K_1\, \sca{\bnbl}{\vc{n}}^2+
K_2\,\sca{\vc{n}}{\bnbl\times\vc{n}}^2
\notag\\
&
+
K_3\,\bigl[\vc{n}\times(\bnbl\times\vc{n})\bigr]^2
\Bigr]\dd v,
\end{align}
\begin{align}
\label{eq:fs-gen}
&
F_s=\frac{1}{2}\, \int_S\Bigl[
W(\vc{n})-K_{24}\,\bigl[
\sca{\bs{\nu}}{\vc{n}}\sca{\bnbl}{\vc{n}}
\notag\\
&
-
\sca{\bs{\nu}}{\sca{\vc{n}}{\bnbl}\vc{n}}
\bigr]
\Bigr]\dd s,
\end{align}
where $K_1$, $K_2$ and $K_3$ are the splay, twist and bend elastic
constants, respectively; $K_{24}$ is the saddle-splay elastic
constant; $\bs{\nu}$ is the outer normal to the surface $S$;
$W(\vc{n})$ is the surface density of the anchoring energy.

An important point is that,
in addition to the anchoring energy which is the anisotropic part of
the surface tension, there is also  the elastic
contribution to the surface free energy that, 
originally, has been indicated
as a part of the elastic energy having the form
of a divergence~\cite{Frank:1958,Luben:pra:1970,Nehr:1971}.
This contribution~---~the so-called saddle-splay term
(the $K_{24}$ term)~---~can generally be viewed as the 
tangential director gradient dependent elastic
part of the surface energy~\cite{Faet:pre1:1994,Perg:pre:2002}.
The other surface elastic term known as the $K_{13}$ term
will not be considered in this paper as it can be
ignored in cases where spatial variations of the density and 
the scalar order parameter are of minor
importance~\cite{Faet:jpfr:1995,Yok:1997,Perg:pre:1999}. 

In the last years $K_{24}$ specific issues have attracted much less
attention than the fundamental difficulties caused by the $K_{13}$
term.  In particular, though the exact measurements of $K_{24}$ are
still missing it was experimentally estimated to be of the order of
the Frank elastic
constants~\cite{Alld:prl:1991,Craw:pra:1992,Pol:pre:1994}.  One of the
most important theoretical results is that the $K_{24}$ term may
induce spontaneous twist deformations in hybrid nematic films with
azimuthally degenerate anchoring conditions~\cite{Perg:pre1:1993}.
Such deformations are manifested in the formation of periodic stripe
domains observed in sufficiently thin hybrid NLC
cells~\cite{Spar:pre:1994,Perg:prl:1994,Lavr:ijmp:1995}.

For planar NLC cells, similar instability of the ground state
in the presence of the $K_{24}$ term was considered in
Refs.~\cite{Kis:mclc:1998,Perg:pre:2000}.
Recently, in Ref.~\cite{Barb:pre3:2002}, Barbero and Pergamenshchik
suggested that in the proximity of the nematic-smectic-$A$ transition 
the $K_{24}$ term grows anomalously large so as to violate
the Ericksen stability conditions~\cite{Eric:1966}:
\begin{equation}
  \label{eq:Erick-orig}
  0< K_{24}< 2\min(K_1,K_2).
\end{equation}
As a result, the uniform 
equilibrium director distribution becomes unstable 
and a periodically modulated nematic phase may occur in sufficiently
thin planar films.

The results of Refs.~\cite{Perg:pre:2000,Barb:pre3:2002,Barb:pre4:2002} are
essentially limited to the special case in which the azimuthal
anchoring strength is identically zero.  But the study of possible
mechanisms leading to the formation of modulated orientational
structures close to the nematic-smectic-$A$ transition requires a
quantitatively accurate description of the instability that goes
beyond this limitation.  At this stage, however, even the instability
scenario as a whole has not been studied in any detail.

In this paper we intend to fill the gap. 
Our primary goal is the comprehensive study
of the instability induced by the $K_{24}$ term in the presence of the
azimuthal anchoring. 

The idea underlying our general theoretical
considerations is that instabilities of this sort occur when
the director fluctuations at confining surfaces become critically divergent.
So, we suggest the method connecting the correlation functions of
director fluctuations and 
the computational procedure applied to perform the stability analysis.
This method is based on separating out the surface part of the
correlator as a correlation function of the fluctuation field induced
by the director fluctuations at confining walls. 

The layout of the paper is as follows.
In Sec.~\ref{sec:corr-funct},
we express the surface part of the static correlation
functions of the director fluctuations as a functional integral over
fluctuations at confining walls and
explicitly relate the procedure for computing 
the correlators to the stability conditions used in our stability
analysis.
 
Analytical results for the planar NLC cell
are described in Sec.~\ref{subsec:corr-fluct-nemat}.
We characterize 
the mirror symmetry properties of the fluctuation harmonics 
and calculate the surface part of the correlator.
We find that the result is a sum of the contributions from 
the two fluctuation modes of different symmetry 
(symmetric and antisymmetric) and derive the stability conditions for
these modes.

Stability of the uniform planar orientational structure is studied 
in Sec.~\ref{sec:stab-pl-str}.
We analyze the parameterization of 
the boundary curve enclosing 
the instability region in the thickness--wavelength
plane and show that, 
in addition to the stability interval~\eqref{eq:Erick-orig},
there are two different intervals for $K_{24}$:
(a)~$2\min(K_1,K_2)<K_{24}< 4K_1K_2/(K_1+K_2)$,
where the critical point is characterized by the critical thickness
$d_c$
and the critical fluctuation wavelength $\lambda_c$;
(b)~$K_{24}> 4K_1K_2/(K_1+K_2)$, where NLC cells of any thickness
are unstable with respect to the short wavelength fluctuations with
$\lambda<\lambda_{\infty}$.
It is found that the critical fluctuation mode is antisymmetric at $K_2<K_1$
and is symmetric at $K_2>K_1$.
The critical thickness and the critical wavelength
are computed as functions of $K_{24}$ and the azimuthal anchoring
strength. We also discuss
the spectrum of director fluctuations at the substrates
near the critical thickness.

Finally, in Sec.~\ref{sec:disc-concl}, we present our results
and make some concluding remarks. Details on some technical results
are relegated to Appendixes~\ref{sec:append} and~\ref{sec:append-b}.

\section{Correlation functions and stability criteria}
\label{sec:corr-funct}

In this section we consider the general procedure for computing the
correlation functions (correlators) of NLC director fluctuations in
confined geometry. 
We remind the reader about the standard approach 
that uses functional integrals to represent
averaging over fluctuations~\cite{Just:bk:1993,Luben:bk:1995}.
In this approach the effect of the confining surface
enters the theory through the boundary conditions for the saddle point
equations (Euler-Lagrange equations) and for the Green functions.

We then describe an alternative procedure, where 
the part of the fluctuation field representing 
the director fluctuations at the surface
is separated out by shifting the integration variable
in the functional integral.
The corresponding part of the correlator
is defined by the
surface part of the free energy~\eqref{eq:fs-gen} and involves
averaging over the fluctuations at the surface.  Finally, we show that
the free energy of these fluctuations determines stability of
orientational structures.

\begin{figure*}[!tbh]
\centering
\resizebox{140mm}{!}{\includegraphics*{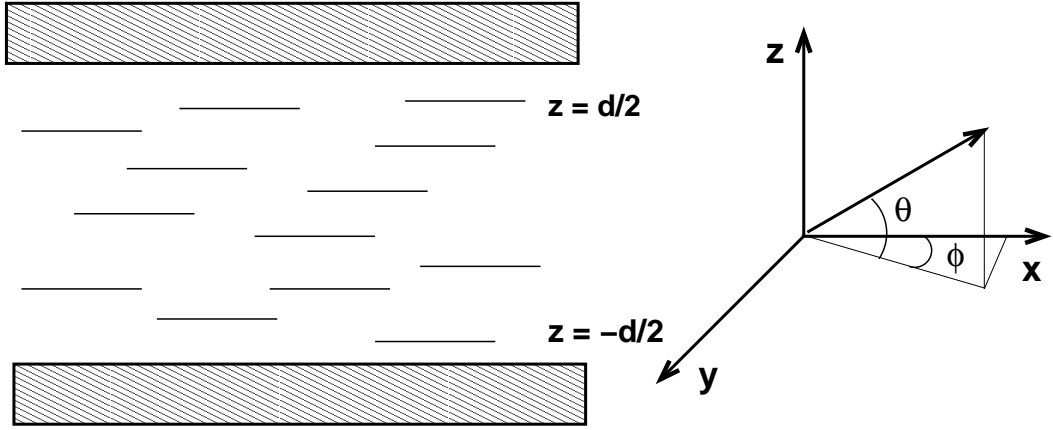}}
\caption{%
Schematic representation of the planar NLC cell.
In-plane and out-of-plane fluctuations are described by the angles
$\phi$ and $\theta$, respectively.
}
\label{fig:nlc-cell}
\end{figure*}

\subsection{Free energy of fluctuations}
\label{subsubsec:free-energy-fluct}

Assuming that 
the director field $\vc{n}_0$ defines
the unperturbed orientational structure,
we shall write the distorted configuration in the form
\begin{align}
  \label{eq:dir-gen}
&
  \vc{n}=\cos\theta\cos\phi\,\vc{n}_0+
\cos\theta\sin\phi\,\vc{n}_1
\notag\\
&
+\sin\theta\,\vc{n}_2,\quad
\sca{\vc{n}_i}{\vc{n}_j}=\delta_{ij},
\end{align}
where the brackets denote the scalar product.
Locally, the vectors $\vc{n}_0$ and $\vc{n}_1$ form 
the $\vc{n}_0$--$\vc{n}_1$ plane, so that
the angles $\phi$ and $\theta$ describe in-plane and out-of-plane
orientational distortions, respectively.

Substituting Eq.~\eqref{eq:dir-gen} into 
Eqs.~\eqref{eq:Fn-gen}~--~\eqref{eq:fs-gen}
will give
the free energy of the director configuration~\eqref{eq:dir-gen}
as a functional of the angles $\phi$ and $\theta$.
For small distortions, $\phi, \theta\ll 1$, 
Eq.~\eqref{eq:dir-gen} reduces to the familiar form
\begin{equation}
  \label{eq:dir-pert}
  \vc{n}\approx\vc{n}_0+\delta\vc{n}_0,\quad
\delta\vc{n}_0=\phi\,\vc{n}_1+\theta\,\vc{n}_2
\end{equation}
and the lowest order approximation
for the free energy of director fluctuations~---~the
so-called Gaussian approximation~---~is given by 
the second order term, $F^{(2)}$, 
of the truncated series expansion for the energy~\eqref{eq:Fn-gen} 
\begin{align}
  \label{eq:happr-1}
&
  F[\vc{n}]\approx F[\vc{n}_0]+F^{(2)}[\bs{\psi}],
\\
  \label{eq:happr-2}
&
  F^{(2)}[\bs{\psi}]=F_{b}^{(2)}[\bs{\psi}]+F_{s}^{(2)}[\bs{\psi}],
\end{align}
where $\bs{\psi}\equiv
\begin{pmatrix}
\psi_1\\
\psi_2
\end{pmatrix}
=
\begin{pmatrix}
\phi\\
\theta
\end{pmatrix}$
is the two-component fluctuation field; $F_{b}^{(2)}$ and $F_{s}^{(2)}$ are
the bulk and the surface parts of
the fluctuation free energy generated by the corresponding terms
of the free energy~\eqref{eq:Fn-gen}.

We can now apply the standard variational procedure to obtain
the saddle point equations for the fluctuation free energy.
(These equations can also be derived as the linearized
Euler-Lagrange equations for the director~\eqref{eq:dir-gen}).
In this section, we only need to specify the general form of the
equations:
\begin{equation}
  \label{eq:EU-lin}
  \frac{\delta F_{b}^{(2)}[\bs{\psi}]}{%
\delta \psi_i(\vc{r})} 
= \hat{K}_{ij} \psi_j(\vc{r})\equiv 
[\hat{K} {\bs{\psi}}(\vc{r})]_i=0,
\end{equation}
where $\hat{K}$ is the matrix differential operator.
Hereafter, matrices and matrix differential operators
will be indicated by hats and
we also assume the summation over repeated indices.

The general structure of the fluctuation free energy can now be
expressed as follows 
\begin{align}
  \label{eq:Fb-repr}
&
F_b^{(2)}[\bs{\psi}]=
\int_V \sca{\bs{\psi}}{\hat{K}\bs{\psi}} \dd v+
\int_S \sca{\bs{\psi}}{\hat{Q}^{(b)}\bs{\psi}} \dd s,
\\
  \label{eq:Fs-repr}
&
F_s^{(2)}[\bs{\psi}]=
\int_S \sca{\bs{\psi}}{\hat{Q}^{(s)}\bs{\psi}} \dd s.
\end{align}
where $\sca{\bs{\varphi}}{\hat{A}\bs{\psi}}\equiv 
\varphi_i \hat{A}_{ij}\psi_j$ and the surface term on the right hand
side of Eq.~\eqref{eq:Fb-repr} results from the integration by parts.
 
There are two most important formal properties
of $F_s^{(2)}$ and $F_b^{(2)}$: 
(a)~director derivatives along the normal to the surface 
do not enter $F_s^{(2)}$;  
and (b)~the bulk free energy $F_b^{(2)}$ is represented by the
symmetric bilinear functional $S_b$, so that
$F_b^{(2)}[\bs{\psi}]=S_b[\bs{\psi},\bs{\psi}]$ and
$S_b[\bs{\varphi},\bs{\psi}]=S_b[\bs{\psi},\bs{\varphi}]$.
In particular, the relation 
\begin{align}
  \label{eq:Green-forml}
&
\int_V \bigl[
\sca{\bs{\varphi}}{\hat{K}\bs{\psi}}- 
\sca{\bs{\psi}}{\hat{K}\bs{\varphi}}
\bigr]\, \dd v
\notag\\
&
=\int_S \bigl[
\sca{\bs{\psi}}{\hat{Q}^{(b)}\bs{\varphi}} - 
\sca{\bs{\varphi}}{\hat{Q}^{(b)}\bs{\psi}}
\bigr]\, \dd s,  
\end{align}
which is a version of the Green formula,
immediately follows from the symmetry of $S_b$.


\begin{figure*}[!thb]
\centering
\resizebox{150mm}{!}{\includegraphics*{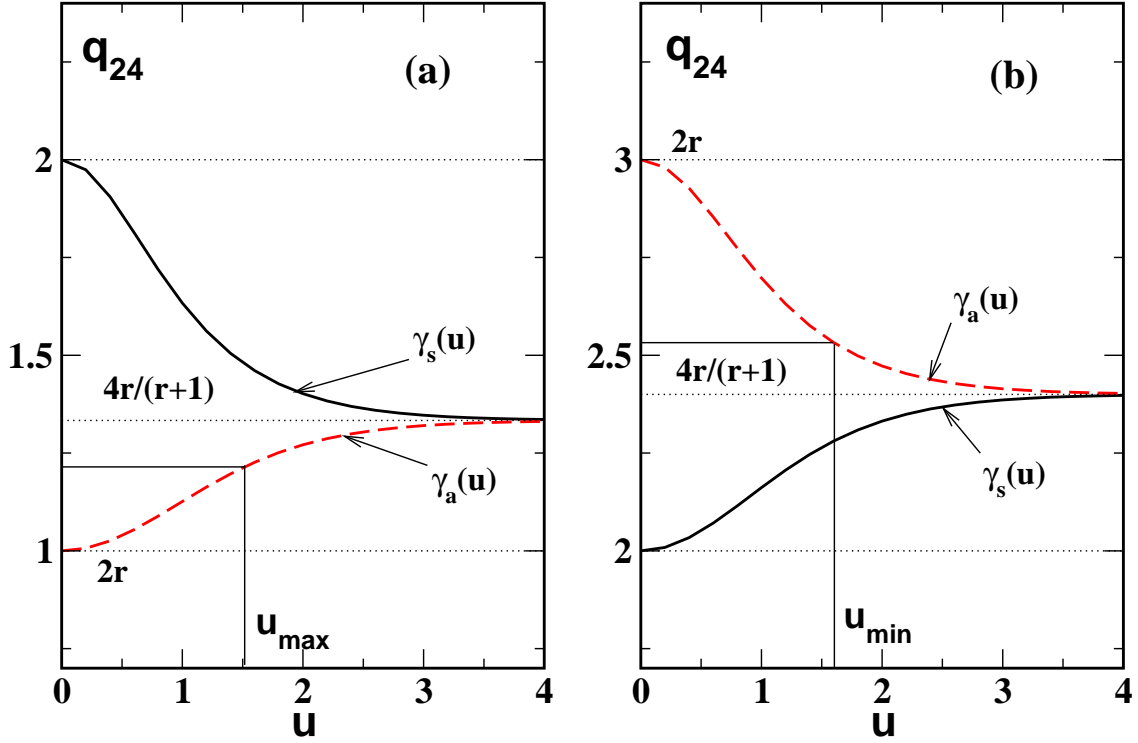}}
\caption{%
The graphs of the functions $\gamma_s(u)$ (solid line) and
$\gamma_a(u)$ (dashed line) in the $u$--$q_{24}$ plane.
Two cases are illustrated: (a)~$r<1$ ($r=0.5$) and 
(b)~$r>1$ ($r=1.5$).
The endpoint of
the instability interval for the critical mode $u<u_{\smx}$
at $q_c^{(1)}<q_{24}<q_c^{(2)}$
is shown on the left.
At $q_c^{(2)}<q_{24}<q_c^{(3)}$, 
the endpoint of
the instability interval for the non-critical mode
$u>u_{\smn}$ is indicated on the right.
}
\label{fig:gamma}
\end{figure*}

\begin{figure*}[!tbh]
\centering
\resizebox{140mm}{!}{\includegraphics*{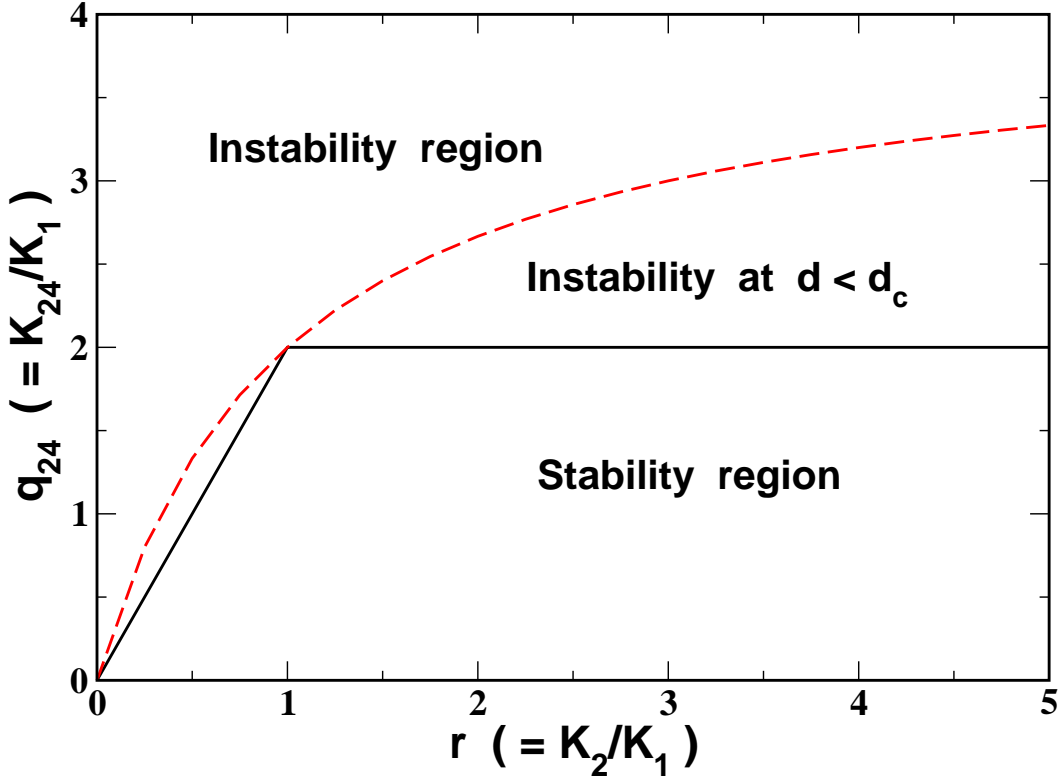}}
\caption{%
Stability diagram in the $r$--$q_{24}$ plane.
There are two critical values of $q_{24}$: 
$q_{c}^{(1)}$ (solid line) and $q_c^{(2)}$ (dashed line).
}
\label{fig:stab}
\end{figure*}

\begin{figure*}[!tbh]
\centering
\resizebox{140mm}{!}{\includegraphics*{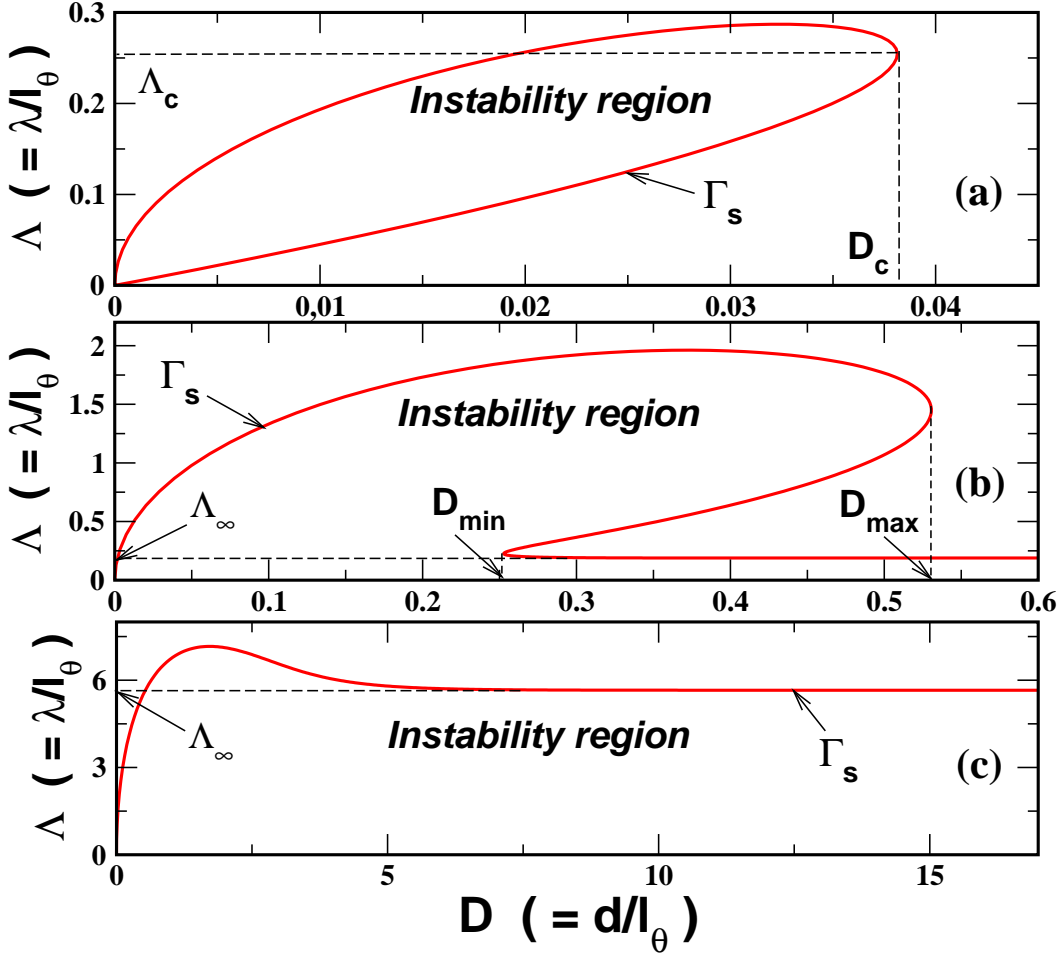}}
\caption{%
Stability diagrams in the $D$--$\Lambda$ plane
at $W_{\phi}=W_{\theta}$ and $r\equiv K_2/K_1=1.5$
for various values of $q_{24}=K_{24}/K_1$:
(a)~$q_c^{(1)}=2<q_{24}=2.1<q_c^{(2)}=2.4$;
(b)~$q_c^{(2)}=2.4<q_{24}=2.43<2r=3$; and
(c)~$2r=3<q_{24}=3.3$.
}
\label{fig:dc-lmbc-1}
\end{figure*}

\begin{figure*}[!tbh]
\centering
\resizebox{140mm}{!}{\includegraphics*{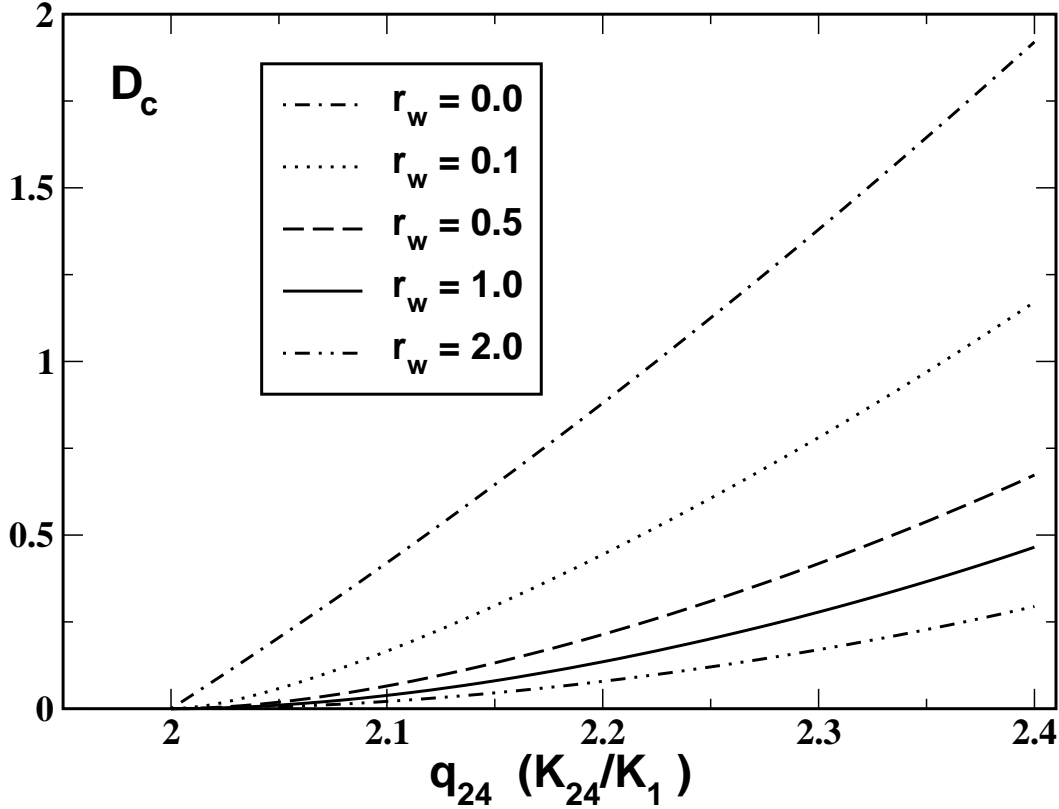}}
\caption{%
Dimensionless parameter of critical thickness 
$D_c$ ($=d_c/l_{\theta}$)
versus 
$q_{24}$ at  $r=1.5$
for various values of 
the azimuthal anchoring parameter $r_w=W_{\phi}/W_{\theta}$.
}
\label{fig:d-q24}
\end{figure*}

\begin{figure*}[!tbh]
\centering
\resizebox{140mm}{!}{\includegraphics*{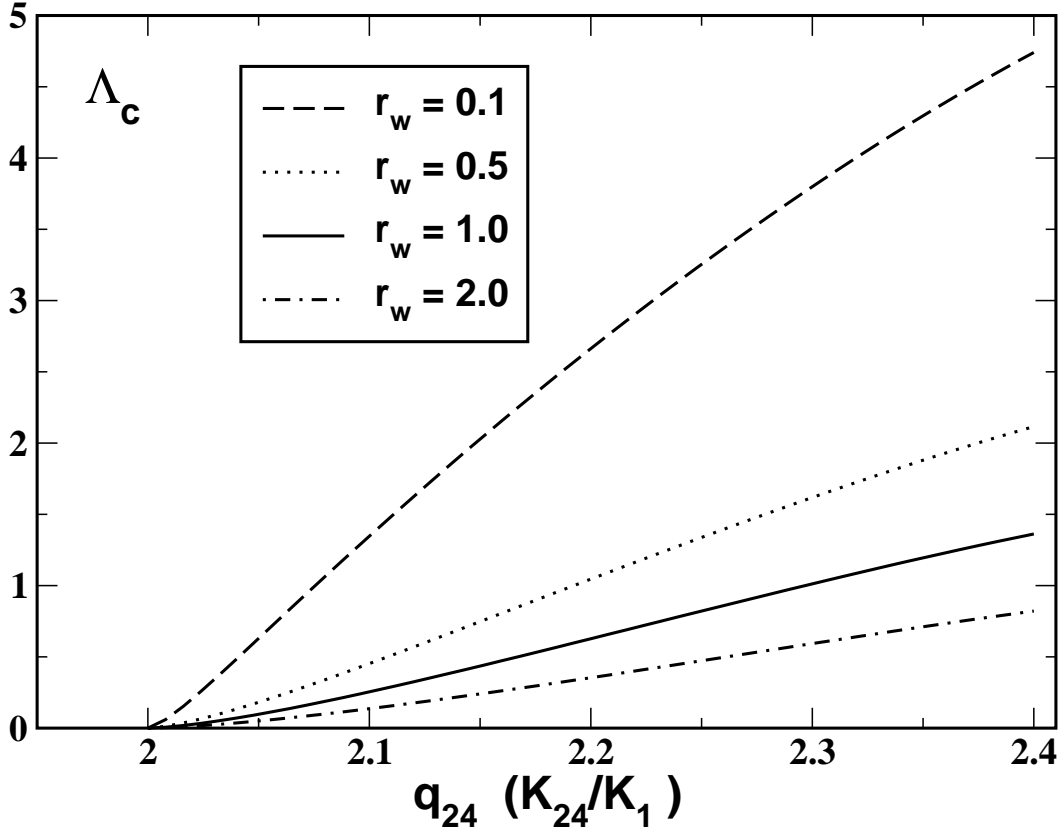}}
\caption{%
Dimensionless parameter of critical wavelength 
$\Lambda_c$ ($=\lambda_c/l_{\theta}$)
versus 
$q_{24}$ at  $r=K_2/K_1=1.5$
for various values of the 
azimuthal anchoring parameter $r_w$.
}
\label{fig:l-q24}
\end{figure*}

\begin{figure*}[!tbh]
\centering
\resizebox{140mm}{!}{\includegraphics*{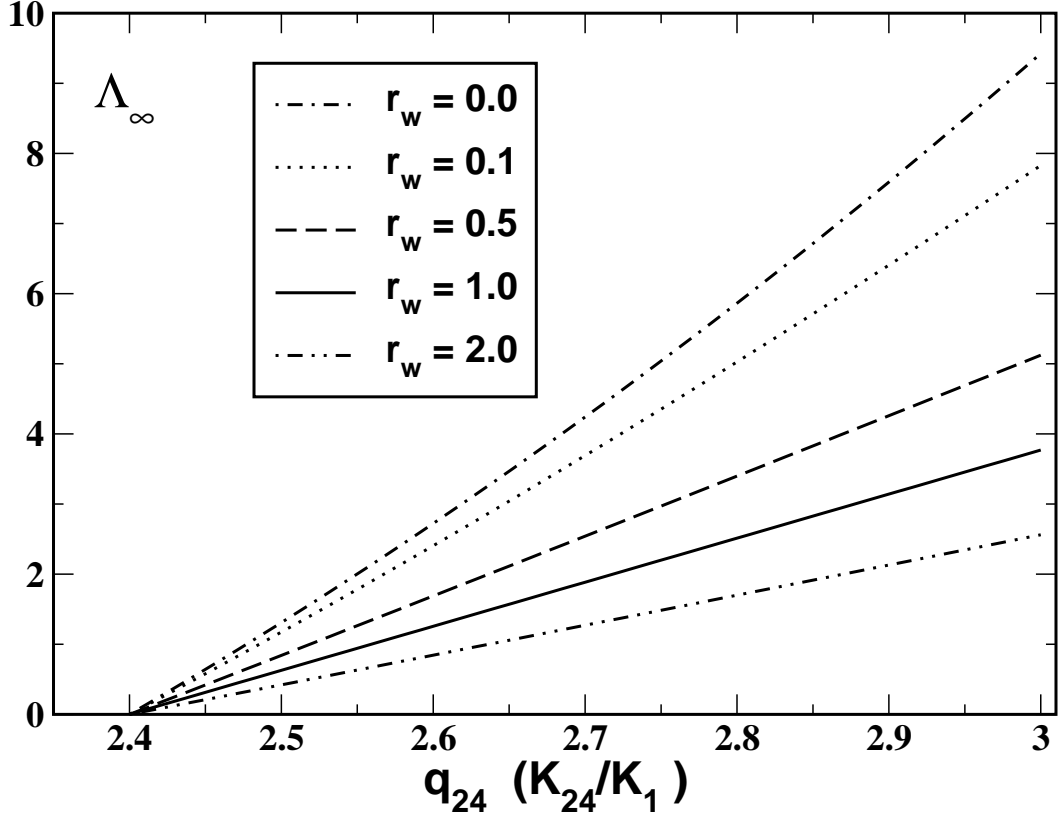}}
\caption{%
Dimensionless parameter of critical asymptotic wavelength 
$\Lambda_{\infty}$ ($=\lambda_{\infty}/l_{\theta}$)
versus 
$q_{24}$ at  $r=K_2/K_1=1.5$
for various values of 
the azimuthal anchoring parameter $r_w$.
}
\label{fig:linf-q24}
\end{figure*}

\begin{figure*}[!tbh]
\centering
\resizebox{140mm}{!}{\includegraphics*{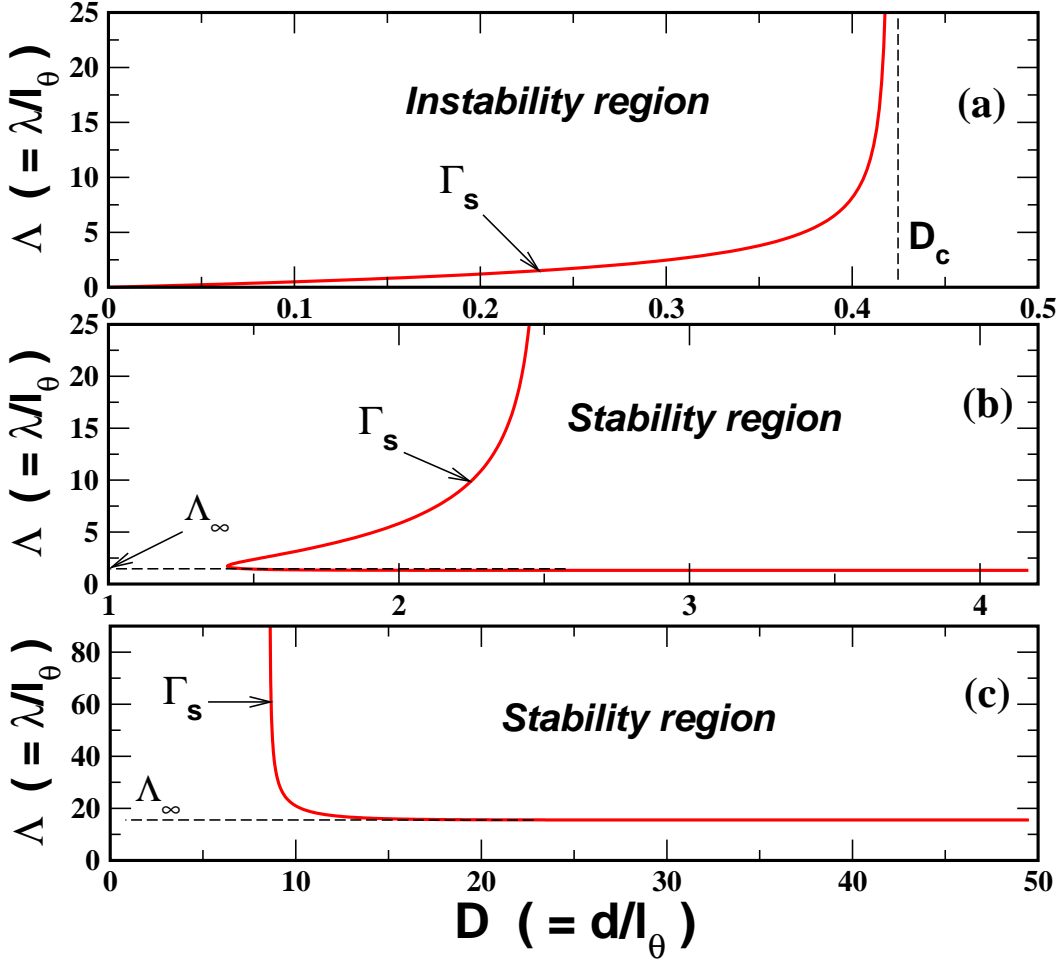}}
\caption{%
Stability diagrams in the $D$--$\Lambda$ plane
at $W_{\phi}=0$ and $r=1.5$
for various values of $q_{24}$:
(a)~$q_c^{(1)}=2<q_{24}=2.1<q_c^{(2)}=2.4$;
(b)~$q_c^{(2)}=2.4<q_{24}=2.5<2r=3$; and
(c)~$2r=3<q_{24}=3.3$.
}
\label{fig:dc-lmbc-0}
\end{figure*}

\begin{figure*}[!tbh]
\centering
\resizebox{140mm}{!}{\includegraphics*{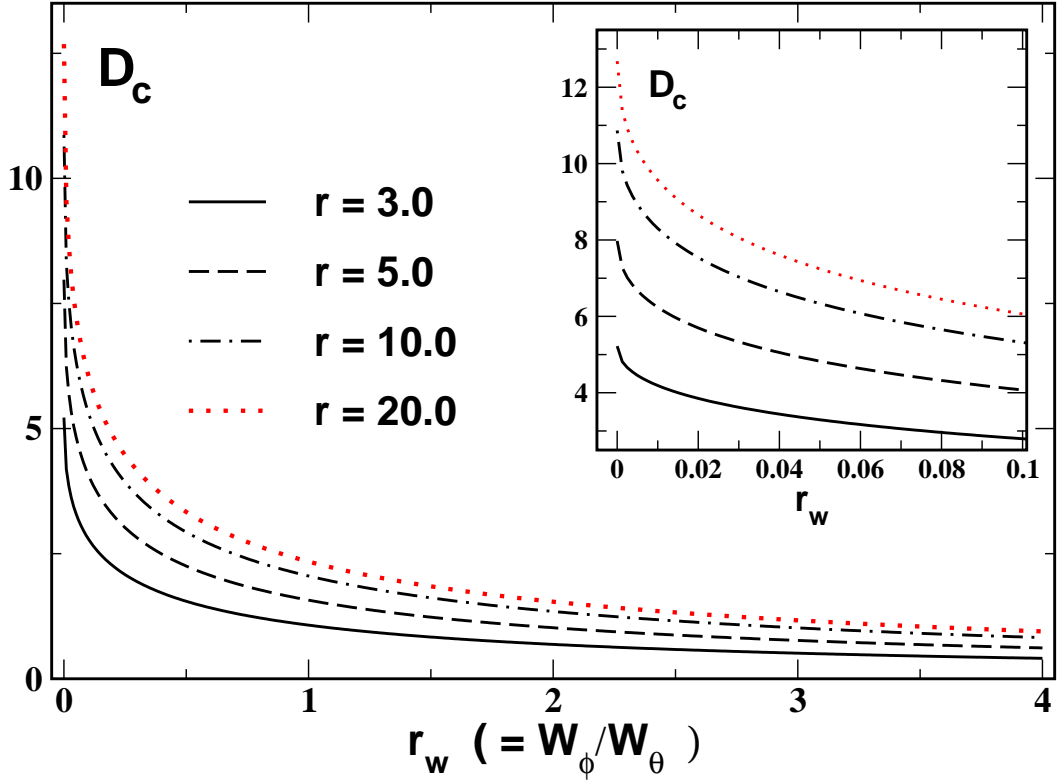}}
\caption{%
Dimensionless parameter of critical thickness 
$D_c$ ($=d_c/l_{\theta}$)
versus
$r_w$ at  $q_{24}=q_c^{(2)}-0.1$
for various values of 
the elatic anisotropy parameter $r$.
Insert at the upper right corner enlarges initial decay in the neighborhood
of the origin.
}
\label{fig:d-rw}
\end{figure*}

\begin{figure*}[!tbh]
\centering
\resizebox{140mm}{!}{\includegraphics*{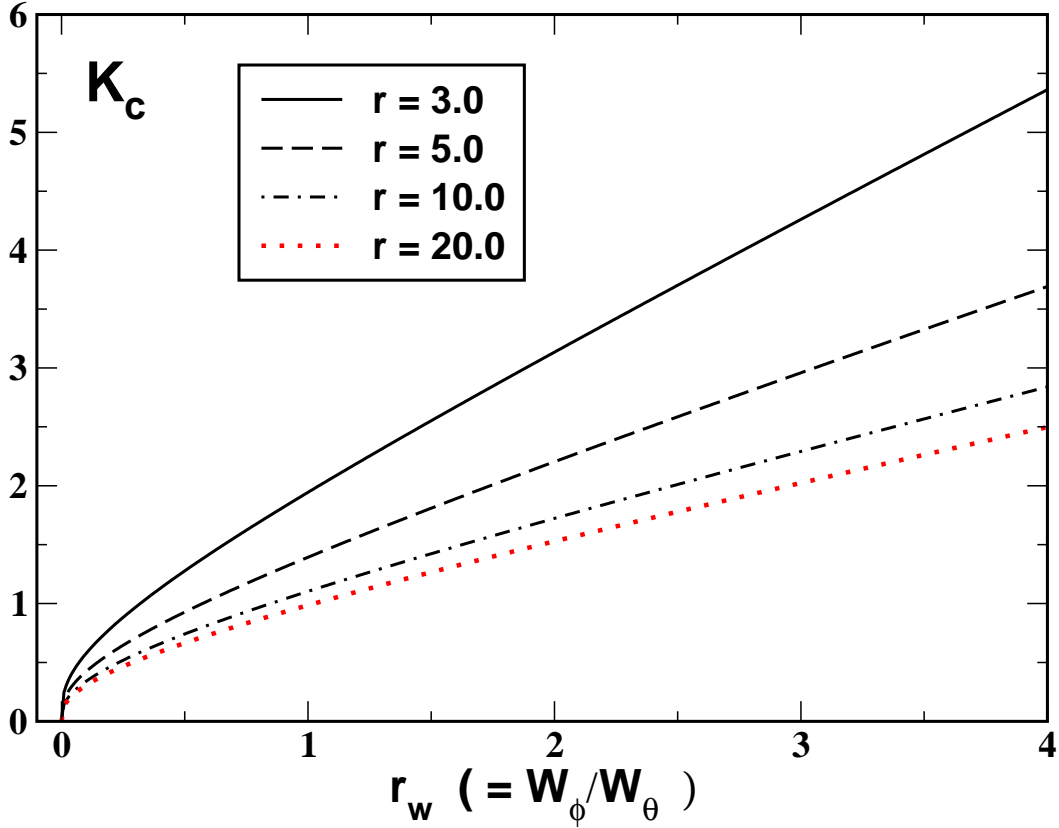}}
\caption{%
Dimensionless parameter of critical wavenumber 
$K_c$ ($=2\pi/\Lambda_c = k_c l_{\theta}$)
versus 
$r_w$ at  $q_{24}=q_c^{(2)}-0.1$
for various values of 
the elastic anisotropy parameter $r$.
}
\label{fig:k-rw}
\end{figure*}

\begin{figure*}[!tbh]
\centering
\resizebox{150mm}{!}{\includegraphics*{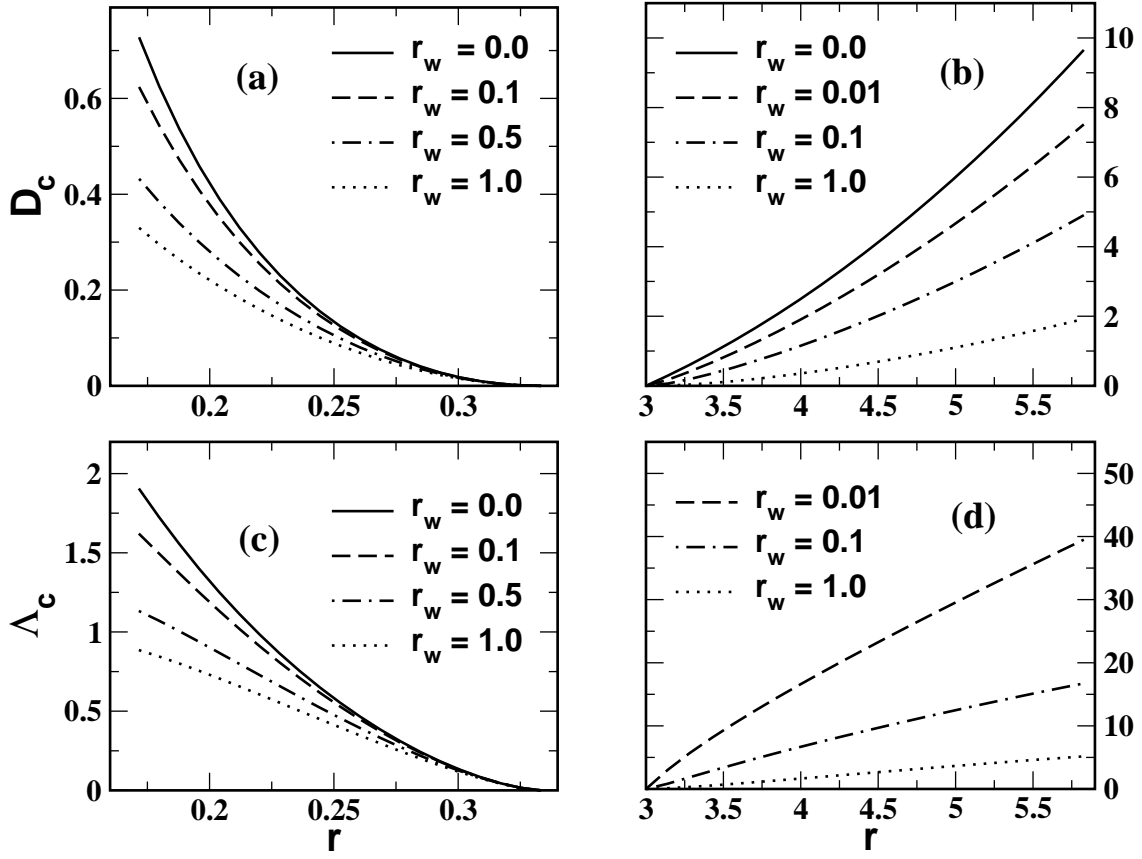}}
\caption{%
(a)-(b)~The critical thickness
and (c)-(d)~the critical wavelength parameters as a function of
the twist-splay ratio $r$ for various values of $r_w$.
The graphs are computed by using the Cauchy relation:
$q_{24}=(1+r)/2$. The saddle-splay parameter $q_{24}$ is
in the instability region~\eqref{eq:part-instb}
when $r$ lies in the intervals:
$3-2\sqrt{2}<r<1/3$ or $3<r<3+2\sqrt{2}$.
}
\label{fig:cauchy}
\end{figure*}

\begin{figure*}[!tbh]
\centering
\resizebox{150mm}{!}{\includegraphics*{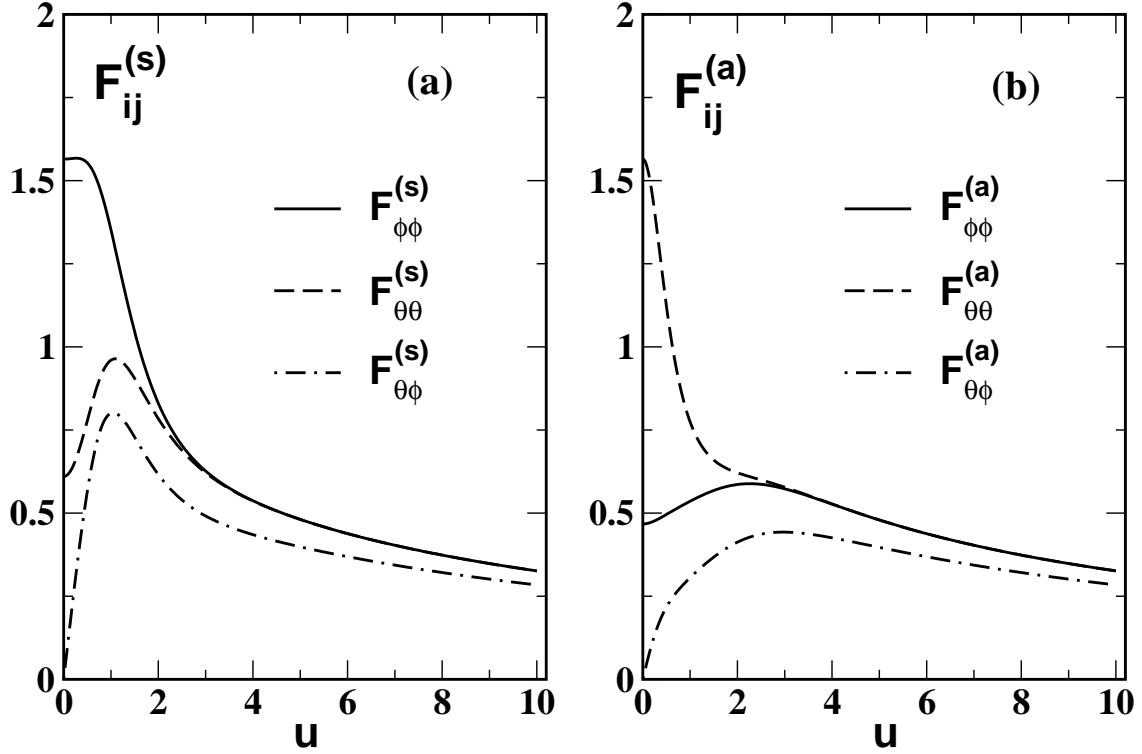}}
\caption{%
Elements of the covariance matrix~\eqref{eq:f-alp} 
as functions of $u$ ($=k_yd/2$)
for (a)~symmetric and (b)~antisymmetric fluctuations
at the lower substrate, $z=-d/2$.
The thickness parameter $D=D_c+1.0=1.27$ 
is  noticeably above its critical value
$D_c=0.27$ at $q_{24}=q_c^{(1)}+0.3=1.7<q_c^{(2)}$,
$r=1.5$ and $r_w=1.0$.
}
\label{fig:corr-1}
\end{figure*}

\begin{figure*}[!tbh]
\centering
\resizebox{150mm}{!}{\includegraphics*{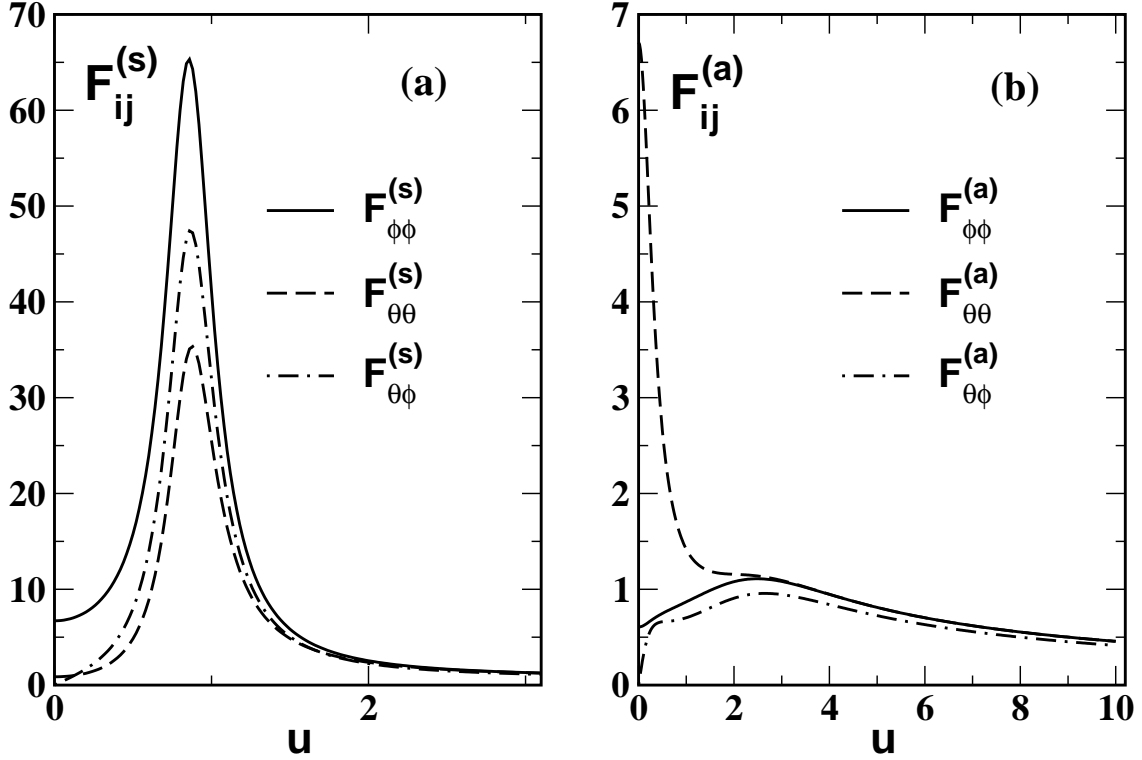}}
\caption{%
Elements of the covariance matrix~\eqref{eq:f-alp}  
as functions of $u$
($=k_yd/2$)
for (a)~symmetric and (b)~antisymmetric fluctuations
at the lower substrate, $z=-d/2$.
The thickness parameter $D=D_c+0.02=0.29$ 
is in the immediate vicinity of its critical value
$D_c=0.27$ at $q_{24}=q_c^{(1)}+0.3=1.7<q_c^{(2)}$,
$r=1.5$ and $r_w=1.0$.
}
\label{fig:corr-02}
\end{figure*}


\subsection{Correlators and Green functions}
\label{subsec:corr-green-funct}

We shall need to write the probability distribution of fluctuations 
at the state of thermal equilibrium in the form
\begin{equation}
  \label{eq:prob-gen}
  P[\bs{\psi}]=Z^{-1}\exp\{-\beta F^{(2)}[\bs{\psi}]\},
\end{equation}
where $\beta\equiv (k_B T)^{-1}$, $k_B$ is the Boltzmann constant,
$T$ is the temperature 
and $Z$ is the partition function given by the functional integral
\begin{equation}
  \label{eq:stsum-gen}
  Z=\int \exp\{-\beta F^{(2)}[\bs{\psi}]\}\DD\bs{\psi},
\end{equation}
where $\DD\bs{\psi}\equiv \DD\psi_1 \DD\psi_2$.
The components of the correlator then be written as 
the averages of the fluctuation field products
\begin{align}
  \label{eq:corr-def}
&
  C_{ij}(\vc{r},\vc{r}')\equiv [\hat{C}(\vc{r},\vc{r}')]_{ij}
=\avr{\psi_i(\vc{r})\psi_j(\vc{r}')}
\notag\\
&
= \int \psi_i(\vc{r})\psi_j(\vc{r}') P[\bs{\psi}]\,\DD\bs{\psi}.
\end{align}
The correlator~\eqref{eq:corr-def} is 
more appropriately known as the 2-point
static correlation function~\cite{Just:bk:1993} 
and so long as we are within the Gaussian
approximation no other correlation functions are required.

For functional integrals, explicit analytical treatment can be rather 
involved~\cite{Glim:bk:1981,Klein:bk:1999}. For our purposes,
it is, however, more convenient to modify the
distribution~\eqref{eq:prob-gen} by adding
a source term to the free energy so as to  
introduce the generating functional
of the correlation functions through the partition
function of the modified distribution~\cite{Just:bk:1993}.  
The free energy is now as follows
\begin{align}
  \label{eq:lc-f2J}
&
  F^{(2)}[\bs{\psi}|\vc{J}]=
  F_{b}^{(2)}[\bs{\psi}|\vc{J}]+ F_{s}^{(2)}[\bs{\psi}]
\notag\\
&
=
F^{(2)}[\bs{\psi}]-\beta^{-1}
\int_V J_i(\vc{r})\psi_i(\vc{r})\dd v,
\end{align}
where $\vc{J}\equiv
\begin{pmatrix}
J_1\\
J_2
\end{pmatrix}$. The relation linking the correlator
and   functional derivatives of the generating
functional is
\begin{align}
  \label{eq:lc-corr-gen-1}
&
C_{ij}(\vc{r},\vc{r}')=
  C_{ij}(\vc{r},\vc{r}'|\vc{J})\bigr\vert_{\vc{J}=0}
\notag\\
&
=
\frac{\delta^2 \ln Z[\vc{J}]}{\delta J_i(\vc{r})\delta J_j(\vc{r}')}
\,\Bigr\vert_{\vc{J}=0}
=
\frac{\delta\bar{\psi}_i(\vc{r}|\vc{J})}{\delta J_j(\vc{r}')}
\,\Bigr\vert_{\vc{J}=0} ,
\end{align}
where
\begin{equation}
  \label{eq:avr-gen}
  \bar{\psi}_i(\vc{r}|\vc{J})=
\int \psi_i(\vc{r})P[\bs{\psi}|\vc{J}]\,\DD\bs{\psi}.
\end{equation}

In our case all the functional integrals are Gaussian and the
saddle point approximation yields the exact results. So, the
averaged fluctuation field~\eqref{eq:avr-gen} 
can be computed as an extremal of 
the free energy~\eqref{eq:lc-f2J} which satisfies the Euler-Lagrange equations
\begin{align}
  \label{eq:avr-eq}
&
\hat{K}_{ij}
  \bar{\psi}_j(\vc{r}|\vc{J})=
\beta^{-1} J_i(\vc{r})
\end{align}
and meets the boundary conditions
\begin{align}
  \label{eq:avr-bc}
&
\hat{Q}_{ij}
  \bar{\psi}_j(\vc{r}|\vc{J})\bigr\vert_{\vc{r}\in S}=0,\quad
\hat{Q}=\hat{Q}^{(b)}+\hat{Q}^{(s)}.
\end{align}

So long as
the boundary value problem
\eqref{eq:avr-eq}--\eqref{eq:avr-bc} 
is linear, the fluctuation field~\eqref{eq:avr-gen}
linearly depend on the source. This relation is given by
\begin{equation}
  \label{eq:avr-sol}
 \bar{\psi}_i(\vc{r}|\vc{J})= \beta^{-1} 
\int_V
G_{ij}(\vc{r},\vc{r}')
 J_j(\vc{r}')\,\dd v', 
\end{equation}
where $\hat{G}$ is the Green function of the operator $\hat{K}$.
Eqs.~\eqref{eq:lc-corr-gen-1}
and~\eqref{eq:avr-sol} show that the correlator is proportional
to the Green function              
\begin{equation}
  \label{eq:Green-corr-rel}
  \hat{C}(\vc{r},\vc{r}')=\beta^{-1} \hat{G}(\vc{r},\vc{r}').
\end{equation}
So, in order compute
the correlator $\hat{C}$ we need to solve
the boundary value problem for the Green function $\hat{G}$:
\begin{subequations}
\label{eq:Green-probl}
\begin{align}
  \label{eq:Green-eq}
&
 \hat{K}_{ik}G_{kj}(\vc{r},\vc{r}')=\delta(\vc{r}-\vc{r}')\delta_{ij},
\\
\label{eq:Green-bc}
&
\hat{Q}_{ik}
  G_{kj}(\vc{r},\vc{r'})\bigr\vert_{\vc{r}\in S}=0, 
\end{align}
\end{subequations}
where $\delta(\vc{r})$ is the delta function.
The problem~\eqref{eq:Green-probl} is at the heart of 
the conventional computational procedures
traditionally used, e.g., in studies of light scattering
in confined liquid 
crystals~\cite{Shal:pre:1993,Valk:ufn:1994,Stal:pre:1996,Whit:prl:1998,Mert:2000}.

The key point is that the effects caused by the anchoring energy
and the surface elasticity constant are solely incorporated into the
boundary conditions~\eqref{eq:Green-bc}. 
These conditions affect eigenfunctions (normal fluctuation modes)
and eigenvalues of the operator $\hat{K}$ that 
can be used to derive the Green function in the form of an expansion over the
eigenfunctions.

The eigenvalues form the spectrum of fluctuations 
and must be positive provided the orientational
structure $\vc{n}_0$ is stable.
Otherwise, the functional integrals~\eqref{eq:stsum-gen}
and~\eqref{eq:corr-def} do not converge. 

We have thus formulated the spectral stability conditions
that turn out to be closely related to
the approach based on the generating functional.
These conditions are in considerable use as stability criteria
in several methods developed to study 
Fr\'{e}edericksz-type transitions in 
confined liquid crystals~\cite{Pikin:bk:1991,Galat:pre:1994}.

\subsection{Surface part of correlator and stability}
\label{subsec:surf-part-corr}

We now pass on to the approach that emphasize the role of the director
fluctuations at the confining surface by using another transformation
of the functional integral~\eqref{eq:corr-def}.  This 
transformation has long been known as an efficient method to perform
Gaussian integrals~\cite{Just:bk:1993} and we, following the general
idea, define the new integration variable $\bs{\varphi}_b$ that
vanishes at the surface by translating the fluctuation field
$\bs{\psi}$:
\begin{equation}
  \label{eq:shift}
  \bs{\psi}=\bs{\varphi}_b+\bs{\varphi},\quad
\bs{\varphi}_b\bigr\vert_{S}=0,
\end{equation}
where
$\bs{\varphi}_b= 
\begin{pmatrix}
\varphi_1^{(b)}\\
\varphi_2^{(b)}
\end{pmatrix}$.

The fluctuation field $\bs{\psi}$ is thus
decomposed into the field, $\bs{\varphi}_b$, 
vanishing at the surface
and the field $\bs{\varphi}$ that accounts
for non-vanishing fluctuations at the surface,
$\bs{\psi}\bigr\vert_{S}=\bs{\varphi}_s$.
When $\bs{\varphi}$ additionally satisfies the Euler-Lagrange
equations~\eqref{eq:EU-lin}
\begin{align}
  \label{eq:shift-eq}
&
  \hat{K}\bs{\varphi}=0,\quad
\bs{\varphi}\bigr\vert_{S}=
\bs{\psi}\bigr\vert_{S}\equiv
\bs{\varphi}_s,
\end{align}
the fluctuations described by $\bs{\varphi}_b$
and $\bs{\varphi}$
will be statistically independent
\begin{align}  
\label{eq:f2-shift}
&
  F^{(2)}[\bs{\varphi}_b+\bs{\varphi}]=
  F_b^{(2)}[\bs{\varphi}_b]+\Phi_s[\bs{\varphi}_s],
\\
  \label{eq:f2-surf}
&
\Phi[\bs{\varphi}_s]=\int_S 
\sca{\bs{\varphi}_s}{\hat{Q}\bs{\varphi}_s}\,\dd s.
\end{align}
Derivation of Eq.~\eqref{eq:f2-shift} relies on the formal
properties listed at the end of Sec.~\ref{subsubsec:free-energy-fluct}.
It involves using the Green formula~\eqref{eq:Green-forml}
and the relation $\hat{Q}^{(s)}\bs{\varphi}_b\bigr\vert_S=0$.

From Eqs.~\eqref{eq:f2-shift} and~\eqref{eq:prob-gen} 
we obtain the probability distributions:
\begin{align}
  \label{eq:prob-bulk}
&
  P_b[\bs{\varphi}_b]=Z_b^{-1}\exp\{-\beta F_b^{(2)}[\bs{\varphi}_b]\},
\\
&
  P_s[\bs{\varphi}_s]=Z_s^{-1}\exp\{-\beta \Phi_s[\bs{\varphi}_s]\},
\label{eq:prob-surf}
\end{align}
where $P_b[\bs{\varphi}_b]$ is the distribution for
the field of fluctuations in the bulk $\bs{\varphi}_b$; and
the distribution $P_s[\bs{\varphi}_s]$ characterizes
the fluctuations at the surface $\bs{\varphi}_s$.

Averaging over the fluctuation fields
$\bs{\varphi}_s$ and $\bs{\varphi}_b$ can now be performed independently.
As a result, we have 
\begin{align}
\label{eq:separ-int}
&
\avr{A_b A_s}=
\int A_b A_s P[\bs{\psi}]\DD\bs{\psi}=
\int A_s P_s[\bs{\varphi}_s]\DD\bs{\varphi}_s
\notag\\
&
\times
\int\limits_{\bs{\varphi}_b\vert_{S}=0}
A_b P_b[\bs{\varphi}_b]\DD\bs{\varphi}_b
=\avr{A_s}_s \avr{A_b}_b,
\end{align}
where $A_s\equiv A_s[\bs{\varphi}_s]$ and $A_b\equiv A_b[\bs{\varphi}_b]$.
 
After
substituting the decomposition~\eqref{eq:shift} into Eq.~\eqref{eq:corr-def} 
and using Eq.~\eqref{eq:separ-int} to carry out averaging
over fluctuations we arrive at the expression for the correlator in
the final form
\begin{equation}
  \label{eq:corr-sum}
  C_{ij}(\vc{r},\vc{r}')=C_{ij}^{(b)}(\vc{r},\vc{r}')+
C_{ij}^{(s)}(\vc{r},\vc{r}').
\end{equation}
The two terms on the right hand side of Eq.~\eqref{eq:corr-sum}
are given by
\begin{align}
  \label{eq:corr-bulk}
&
 C_{ij}^{(b)}(\vc{r},\vc{r}')=
\avr{\varphi_i^{(b)}(\vc{r})\varphi_j^{(b)}(\vc{r}')}_{b}, 
\\
  \label{eq:corr-surf}
&
 C_{ij}^{(s)}(\vc{r},\vc{r}')=
\avr{\varphi_i(\vc{r}|\bs{\varphi}_s)
\varphi_j(\vc{r}'|\bs{\varphi}_s)}_{s},
\end{align}
where notations for the argument of $\bs{\varphi}$
indicate the  dependence of the field $\bs{\varphi}$ on
$\bs{\varphi}_s$  (see Eq.~\eqref{eq:shift-eq}). 

The correlator $\hat{C}^{(b)}$ 
is entirely determined by the bulk director fluctuations
with the probability distribution~\eqref{eq:prob-bulk}. 
In this case the
fluctuations at the surface are suppressed and
the boundary conditions for $\hat{C}^{(b)}$ are
\begin{equation}
  \label{eq:str-bc}
  C_{ij}^{(b)}(\vc{r},\vc{r}')\bigr\vert_{\vc{r}\in S}=0.
\end{equation}
These conditions can be described as the limit
of infinitely strong anchoring for Eq.~\eqref{eq:Green-bc}.
The result is that $\hat{C}=\hat{C}^{(b)}$ in
the case of strong anchoring.

When the anchoring is not infinitely strong, 
the boundary conditions~\eqref{eq:Green-bc}
differ from the strong anchoring conditions~\eqref{eq:str-bc}.
Now the fluctuations at the confining wall have not been suppressed
completely and are characterized by the probability
distribution~\eqref{eq:prob-surf} with the free
energy~\eqref{eq:f2-surf}.
Eq.~\eqref{eq:shift-eq} shows that
these fluctuations transmitted into the bulk give rise to 
the fluctuation field $\bs{\varphi}$.
Eq.~\eqref{eq:corr-surf} gives
the correlator of the field $\bs{\varphi}$ induced by the fluctuations
at the surface and determines the
difference between $\hat{C}$ and $\hat{C}^{(b)}$. 
In what follows the correlator $\hat{C}^{(s)}$ will be referred to as the
surface part of the correlator $\hat{C}$.  

It is of particular interest that the surface part of the
correlator is written as an average with the
distribution~\eqref{eq:prob-surf}.
When  the free energy~\eqref{eq:f2-surf}
is not positive definite and the conditions
\begin{equation}
  \label{eq:stab-crit}
  \sca{\bs{\varphi}_s}{\hat{Q}\bs{\varphi}_s}>0
\end{equation}
are violated, the correlator~\eqref{eq:corr-surf} becomes divergent. 
So, stability of the orientational structure with
respect to the director fluctuations at the confining surface
is determined by the stability conditions~\eqref{eq:stab-crit}.
Instabilities induced by such fluctuations are of our primary concern,
since neither the anchoring energy nor the $K_{24}$ term may affect the
correlator of the bulk fluctuation field~\eqref{eq:corr-bulk}.

Technically, using the conditions~\eqref{eq:stab-crit} enormously
simplifies stability analysis as compared to exploring the spectrum of
fluctuations~\cite{Galat:pre:1994}. The procedure involves two steps:
(a)~solving the boundary value problem for the Euler-Lagrange
equations~\eqref{eq:shift-eq}; and (b)~evaluating the free energy of
the fluctuations at the surface~\eqref{eq:f2-surf}.
Finally, the stability conditions are derived as the conditions
for the free energy to be positive definite.  

Equivalent computational schemes have
already been used to study surface elasticity effects
in Refs.~\cite{Kis:jetp:1995,Kis:mclc:1998,Barb:pre4:2002}.
We have thus developed the method at the very least linking these schemes
and the surface part of the correlator that takes into account
the director fluctuations at the confining surface.

In the next section we apply the above described procedure to the case
of the planar orientational structure in NLC cell.  As an advantage of
our approach, the study of the fluctuations at the bounding plates
will be naturally incorporated into the stability analysis

\section{Director fluctuations in nematic cell}
\label{subsec:corr-fluct-nemat}

In this section we consider a NLC planar cell of the thickness $d$
sandwiched between two substrates that are both
normal to the $z$ axis: $z=-d/2$ and $z=d/2$.
Anchoring conditions at both substrates are planar with the vector of
the easy orientation directed along the $x$ axis.
Thus, the uniform planar director distribution $\vc{n}_0=\vc{e}_x$ 
gives the undistorted orientational structure.

The distorted director~\eqref{eq:dir-gen} with $\vc{n}_1=\vc{e}_y$
and $\vc{n}_2=\vc{e}_z$ is characterized by  
the angles $\phi$ and $\theta$ shown in Fig.~\ref{fig:nlc-cell}.
For small angles, the approximate second order expression for the anchoring
potential is
\begin{equation}
  \label{eq:anchor}
  W(\vc{n})\approx W(\vc{n}_0)+
W_{\theta}\theta^2+ W_{\phi}\phi^2,
\end{equation}
where $W_{\theta}$ and $W_{\phi}$ are the zenithal (polar) and the azimuthal
anchoring strengths.
Note that Eq.~\eqref{eq:anchor} does not imply using  
the Rapini-Papoular potential~\cite{Rap:1969}, where
$W_{\theta}=W_{\phi}$.
It is, however, more reasonable to consider the case in which
the zenithal and the azimuthal anchoring energies differ,
because the surface energy costs for splay-bend and twist
director deformations in interfacial layers 
are different (recent discussion can be found in
Ref.~\cite{Zhao:2002}).

The cell is invariant under translations in the $x$--$y$ plane
and we impose the periodic conditions on the angles:
$\bs{\psi}(x+L_x,y,z)=\bs{\psi}(x,y,z)$ and
$\bs{\psi}(x,y+L_y,z)=\bs{\psi}(x,y,z)$, where $L_x$ and $L_y$
are the characteristic lengths of the cell along the $x$ and $y$
axes, respectively. We can now write down the Fourier series expansion for
the fluctuations:
\begin{align}
  \label{eq:pl-Four}
&
  \bs{\psi}=\sum_{\vc{m},\,m_y\ge 0}
\bigl\{
\bs{\psi}_{\vc{m}}(z)\exp[i(k_x x+k_y y)]
\notag\\
&
+\cnj{\bs{\psi}}_{\vc{m}}(z)\exp[-i(k_x x +k_y y)]
\bigr\},
\end{align}
where $\vc{m}=(m_x,m_y)$, $k_x=2\pi m_x/L_x$,
$k_y=2\pi m_y/L_y$ and 
an asterisk indicate complex conjugation.

As is shown in Appendix~\ref{sec:append}, 
the fluctuation harmonics with non-vanishing
$k_x$ do not produce additional instability.
So, we shall restrict ourselves to the case in which $k_x=0$.
Owing to the translational symmetry,
the fluctuation harmonics are statistically independent
in the Gaussian approximation 
and the free energy of fluctuations takes the form of 
a sum of the energies of the fluctuation modes: 
\begin{equation}
  \label{eq:FFm}
  F^{(2)}[\bs{\psi}]= \sum_{m\ge 0} 
(2-\delta_{m,0})\,
F_{m}^{(2)}[\bs{\psi}_{m}],
\end{equation}
where $m\equiv m_y$.

Calculation of the free energy is rather straightforward
(some details and related comments are given Appendix~\ref{sec:append})
and we present the result in matrix notations
for the modified fluctuation harmonics:  
\begin{equation}
  \label{eq:notat-1}
  \bs{\psi}_{m} \to
\bs{\psi}=
\begin{pmatrix}
1 & 0\\
0 & i
\end{pmatrix}
\bs{\psi}_{m}.
\end{equation}
For brevity, we drop the index $m$ from notations for
the fluctuation field.
The energy then can be written in the form
\begin{equation}
  \label{eq:F2m}
 F_{m}^{(2)}[\bs{\psi}]
=\frac{K_1 S}{d}\,
\bigl\{\,S_{m}^{(b)}[\bs{\psi}]+S_m^{(s)}[\bs{\psi}]\,\bigr\}, 
\end{equation}
\begin{align}
  \label{eq:Smb}
&
  S_{m}^{(b)}[\bs{\psi}]=u\,\int_{-u}^{u}\dd\tau
\Bigl[
\hcnj{[\prt{\tau}\bs{\psi}]}\hat{A}\,\prt{\tau}\bs{\psi}
+\frac{1}{2}\,\hcnj{\bs{\psi}}\hat{B}\prt{\tau}\bs{\psi}
\notag\\
&
+\frac{1}{2}
\hcnj{[\prt{\tau}\bs{\psi}]}\hcnj{\hat{B}}\bs{\psi}
+ \hcnj{\bs{\psi}}\hat{C}\bs{\psi}
\Bigr],
\end{align}

\begin{align}
\label{eq:Sms}
&
S_{m}^{(s)}[\bs{\psi}]=\sum_{\mu=\pm 1}
\hcnj{\bs{\psi}}\hat{Q}_{\mu}^{(s)}\bs{\psi}\Bigr\vert_{\tau=\mu u},
\end{align}

\begin{align}
  \label{eq:AC-def}
&
  \hat{A}=\begin{pmatrix} r & 0\\ 0 & 1\end{pmatrix},
\quad
\hat{C}=\begin{pmatrix} 1 & 0\\ 0 & r\end{pmatrix},
\end{align}
\begin{align}
&
\hat{B}=(r-1)\begin{pmatrix} 0 & 1\\ -1 & 0\end{pmatrix},
  \label{eq:B-def}
\end{align}
\begin{align}
\label{eq:Q-mu-def}
&
\hat{Q}_{\mu}^{(s)}=\mu u (q_{24}-(1+r)/2)
\begin{pmatrix} 0 & 1\\ 1 & 0\end{pmatrix}
\notag\\
&
+
\begin{pmatrix} w_{\phi} & 0\\ 0 & w_{\theta}\end{pmatrix},
\end{align}
where $S=L_x L_y$ is the area of the substrates;  
$\prt{\tau}\equiv\dfrac{\partial}{\partial\tau}$ and
a cross indicate Hermitian conjugation.
The dimensionless parameters used in 
Eqs.~\eqref{eq:Smb}--\eqref{eq:B-def} are:
\begin{align}
  \label{eq:notat-rq}
&
  \tau=k_y z,\;  r=\frac{K_2}{K_1},\; 
q_{24}=\frac{K_{24}}{K_1},
\\
&
u=\frac{k_y d}{2},\;
w_{\phi,\theta}=\frac{W_{\phi,\theta} d}{2 K_1}=
\frac{d}{2l_{\phi,\theta}},
\label{eq:notat-w}
\end{align}
where $l_{\theta}$ and $l_{\phi}$ 
are the zenithal and the azimuthal anchoring extrapolation lengths,
respectively.

\subsection{Mirror symmetry and parity of fluctuations}
\label{subsec:mirr-symm}

In order to evaluate the correlator of director fluctuations and
study the stability of the planar structure, $\vc{n}_0=\vc{e}_x$,
we shall need to solve the Euler-Lagrange equations
for the fluctuation free energy functional~\eqref{eq:F2m}: 
 \begin{align}
   \label{eq:EU-Smb}
 &
   \hat{L}\bs{\varphi}=0,
 \\
  \label{eq:L-Smb}
  &
  \hat{L}=\hat{A}\prt{\tau}^2-\hat{B}\prt{\tau}-\hat{C}.
\end{align}
These equations
are invariant under the mirror symmetry transformation~\cite{Kis:mclc:1998}:
\begin{equation}
  \label{eq:P-transf}
  \bs{\varphi}(\tau)\rightarrow\hat{P}\,\bs{\varphi}(-\tau),\:
\hat{P}=\begin{pmatrix}1 & 0\\ 0 & -1\end{pmatrix}
\end{equation}
Algebraically, this result follows because the matrices
$\hat{P}$, $\hat{A}$ and $\hat{C}$ are commuting,
$\hat{P}\hat{A}-\hat{A}\hat{P}=\hat{P}\hat{C}-\hat{C}\hat{P}=0$,
whereas the matrices $\hat{P}$ and $\hat{B}$ anticommute,
$\hat{P}\hat{B}+\hat{B}\hat{P}=0$.

The identity $\hat{P}^{2}=\hat{I}$,
where $\hat{I}$ is the identity matrix, 
shows that invariant
sets of the solutions of
Eq.~\eqref{eq:EU-Smb} can be characterized by the parity
with respect to the transformation~\eqref{eq:P-transf}: 
\begin{equation}
  \label{eq:parity}
  \bs{\psi}_{s,a}(-\tau)=\pm\hat{P} \bs{\psi}_{s,a}(\tau),
\end{equation}
where $\bs{\psi}_s$ and $\bs{\psi}_a$ will be referred to as 
the symmetric and the antisymmetric fluctuation modes, respectively.
For symmetric [antisymmetric] 
fluctuation fields, the parity relation~\eqref{eq:parity} 
means that the in-plane and 
the out-of-plane components, $\phi$ and $\theta$, 
are represented by even [odd] and odd
[even] functions of $\tau$, correspondingly:
$\phi(-\tau)=\phi(\tau)$ [$\phi(-\tau)=-\phi(\tau)$]
and 
$\theta(-\tau)=-\theta(\tau)$ [$\theta(-\tau)=\theta(\tau)$].

The general solution is now a sum of the symmetric
and the antisymmetric modes:
\begin{align}
  \label{eq:gen-sol}
&
  \bs{\varphi}(\tau)=\bs{\psi}_s(\tau)+\bs{\psi}_a(\tau).
\end{align}
We shall write expressions for the modes in matrix notations
through the two fundamental matrices
$\hat{\Psi}^{(s)}(\tau)$ and $\hat{\Psi}^{(a)}(\tau)$
composed from solutions of the corresponding symmetry.
These $2\times 2$ matrices satisfy both the Euler-Lagrange
equation~\eqref{eq:EU-Smb} and the parity relation~\eqref{eq:parity}.  
In addition, $\hat{\Psi}^{(\alpha)}(\tau)$
will be conveniently normalized by the condition:
$\hat{\Psi}^{(\alpha)}(u)=\hat{I}$.
It can be verified that solving Eq.~\eqref{eq:EU-Smb} yields 
the following result:
\begin{align}
  \label{eq:sol-sa-1}
&
\bs{\psi}_{\alpha}(\tau)=\hat{\Psi}^{(\alpha)}(\tau)
\bs{\varphi}_{\alpha}, \quad \alpha= s, a,
\\
\label{eq:sol-sa-2}
&
\hat{\Psi}^{(\alpha)}(\tau)=\hat{\Phi}_\alpha(\tau)
[\hat{\Phi}_\alpha(u)]^{-1},
\end{align}
\begin{widetext}
\begin{subequations}
\label{eq:Phi_alpha}
\begin{align}
  \label{eq:Phi-s}
&
  \hat{\Phi}_s(\tau)=
\begin{pmatrix}
\cosh\tau+\rho\tau\sinh\tau & -\rho\tau\sinh\tau\\
\rho\tau\cosh\tau & \sinh\tau - \rho\tau\cosh\tau
\end{pmatrix},
\\
  \label{eq:Phi-a}
&  
\hat{\Phi}_a(\tau)=
\begin{pmatrix}
\sinh\tau+\rho\tau\cosh\tau & -\rho\tau\cosh\tau\\
\rho\tau\sinh\tau & \cosh\tau - \rho\tau\sinh\tau
\end{pmatrix},
\end{align}
\end{subequations}
\end{widetext} 
where
$\bs{\varphi}_{\alpha}$ are the vectors of integration constants
and
\begin{equation}
  \label{eq:rho}
\rho=(1-r)/(1+r).  
\end{equation}

After substituting Eqs.~\eqref{eq:gen-sol}--\eqref{eq:sol-sa-1} 
into Eq.~\eqref{eq:F2m} and using the symmetry
relation~\eqref{eq:parity},
we find that the symmetric and the antisymmetric fluctuations
independently contribute to
the free energy of the fluctuation field~\eqref{eq:gen-sol}:
\begin{align}
  \label{eq:en-s-a}
&  
F_{m}^{(2)}[\bs{\varphi}]
=\frac{2K_1 S}{d}\,
\bigl(\hcnj{\bs{\varphi}}_s\hat{M}_s\bs{\varphi}_s
+\hcnj{\bs{\varphi}}_a\hat{M}_a\bs{\varphi}_a
\bigr),
\\
\label{eq:gen-M-sa}
&
\hat{M}_{\alpha}=
u\,\Bigl(
\hat{A}\prt{\tau}\hat{\Psi}^{(\alpha)}\bigr\vert_{\tau=u}
-\frac{1}{2}\hat{B}
\Bigr)+\hat{Q}_{+}^{(s)}.
\end{align}
We can now combine Eqs.~\eqref{eq:AC-def}--\eqref{eq:Q-mu-def}
and Eqs.~\eqref{eq:sol-sa-2}--\eqref{eq:rho}
to derive expressions for the matrices $\hat{M}_s$ and $\hat{M}_a$
from Eq.~\eqref{eq:gen-M-sa}. The result is given by
\begin{widetext}
\begin{subequations}
\label{eq:M_alpha}
\begin{align}
  \label{eq:M_s}
&
  \hat{M}_s=\frac{1}{\beta_s}
\begin{pmatrix}
(1-\rho) u \tanh^2 u +w_{\phi}\beta_s &
u(q_{24}\beta_s-(1-\rho)\tanh u)\\
u(q_{24}\beta_s-(1-\rho)\tanh u)&
(1-\rho) u  +w_{\theta}\beta_s 
\end{pmatrix},
\\
  \label{eq:M_a}
&
  \hat{M}_a=\frac{1}{\beta_a}
\begin{pmatrix}
(1-\rho) u  +w_{\phi}\beta_a &
u(q_{24}\beta_a-(1-\rho)\tanh u)\\
u(q_{24}\beta_a-(1-\rho)\tanh u)&
(1-\rho) u \tanh^2 u  +w_{\theta}\beta_a 
\end{pmatrix},
\end{align}
\end{subequations}
\end{widetext}
where
\begin{equation}
\label{eq:bet-a-s}
\beta_{s,a}=\tanh u\mp\rho u (1-\tanh^2 u).
\end{equation}

\subsection{Director fluctuations at substrates}
\label{sec:direct-fluct-subs}

Now we apply the analytical results of
Sec.~\ref{sec:direct-fluct-subs}
to evaluate the energy of fluctuation 
harmonics at the substrates
and the surface part of the correlation function.
According to the procedure described in
Sec.~\ref{subsec:surf-part-corr},
the first step is to solve the boundary value
problem~\eqref{eq:shift-eq}.

To this end we shall find the fluctuation field that
satisfies the Euler-Lagrange equations~\eqref{eq:EU-Smb}
and the boundary conditions
\begin{align}
  \label{eq:bc-varphi}
  \bs{\varphi}(\pm u)=\bs{\varphi}^{(\pm)},
\end{align}
where $\bs{\varphi}^{(+)}$ [$\bs{\varphi}^{(-)}$]
is the value of the fluctuation field at the upper [lower]
substrate of the cell.
By analogy with the representation
given in Eqs.~\eqref{eq:gen-sol} and~\eqref{eq:sol-sa-1},
we can write the field in the form 
\begin{align}
\label{eq:sol-bc}
\bs{\varphi}(\tau)=
\hat{\Psi}^{(+)}(\tau)\bs{\varphi}^{(+)}+
\hat{\Psi}^{(-)}(\tau)\bs{\varphi}^{(-)},
\\
\label{eq:Psi-pm-bc}
\hat{\Psi}^{(\mu)}(\nu u)=\delta_{\mu\nu}\hat{I},
\quad \mu, \nu =\pm 1,
\end{align}
where the matrices
$\hat{\Psi}^{(+)}$ and $\hat{\Psi}^{(-)}$ 
are the solutions of the Euler-Lagrange equation~\eqref{eq:EU-Smb}
that are equal to the identity
matrix at one substrate and vanish at the other.

It can be readily verified that
$\hat{\Psi}^{(+)}$ and $\hat{\Psi}^{(-)}$
are related to
the matrices derived in Eqs.~\eqref{eq:sol-sa-2}--\eqref{eq:Phi-a}
through the linear combinations
\begin{subequations}
\label{eq:Psi-pm}
\begin{align}
  \label{eq:Psi-p}
&
 \hat{\Psi}^{(+)}=\frac{1}{2}\,\bigl\{
\hat{\Psi}^{(s)} + \hat{\Psi}^{(a)}
\bigr\},
\\
  \label{eq:Psi-m}
&
 \hat{\Psi}^{(-)}=\frac{1}{2}\,\bigl\{
\hat{\Psi}^{(s)} - \hat{\Psi}^{(a)}
\bigr\} \hat{P}.
\end{align}
\end{subequations}
The fluctuation field~\eqref{eq:sol-bc} thus has been defined.

Clearly, Eq.~\eqref{eq:sol-bc} and Eq.~\eqref{eq:gen-sol} present the
fluctuation field $\bs{\varphi}$ in two different forms.
This in combination with Eqs.~\eqref{eq:Psi-p}--\eqref{eq:Psi-m} 
yields the relation between
the vectors $\bs{\varphi}^{(\pm)}$
and $\bs{\varphi}_{s,a}$:
\begin{align}
  \label{eq:pm-to-as}
\bs{\phi}\equiv
  \begin{pmatrix} \bs{\varphi}^{(+)} \\
    \bs{\varphi}^{(-)}\end{pmatrix}
= \hat{T}
  \begin{pmatrix} \bs{\varphi}_{s} \\ \bs{\varphi}_{a}\end{pmatrix},
\end{align}
where 
$ 
\hat{T}=\begin{pmatrix}
\hat{I} & \hat{I} \\
\hat{P} & -\hat{P} 
\end{pmatrix}
$.

The general expression for
the energy of fluctuations at the surfaces has been derived in
Eq.~\eqref{eq:f2-surf}. In our case this energy 
is a quadratic form of the
boundary values of the fluctuation field~\eqref{eq:sol-bc},
$\bs{\varphi}^{(+)}$ and $\bs{\varphi}^{(-)}$,
that can be evaluated
by substituting Eq.~\eqref{eq:sol-bc} into Eq.~\eqref{eq:F2m}.
From the other hand, the energy expressed in terms of 
$\bs{\varphi}_{s}$ and $\bs{\varphi}_{a}$
has already been obtained in Eq.~\eqref{eq:en-s-a}.
So, with the relation~\eqref{eq:pm-to-as} we are left with
the following result
\begin{align}
  \label{eq:en-surf}
&
\Phi^{(s)}_{m}(\bs{\varphi}^{(+)},\bs{\varphi}^{(-)})
=  F_{m}^{(2)}[\bs{\varphi}]
\notag\\
&
=\frac{K_1 S}{d}\,
\hcnj{\bs{\phi}}\hat{N}\bs{\phi},
\\
&
\label{eq:N}
\hcnj{\hat{T}}\hat{N}\hat{T}= 2
\begin{pmatrix}
\hat{M}_s & 0\\
0 & \hat{M}_a
\end{pmatrix},
\end{align}
where the matrix $\hat{N}$ is described in 
Appendix~\ref{sec:append-b}
(see Eqs.~\eqref{eq:N-tot-b}--\eqref{eq:N-b-b}). 

After substituting the fluctuation field into the expression
for the surface part of the correlator~\eqref{eq:corr-surf}
we have
\begin{align}
  \label{eq:avr-surf}
&
C_{ij}^{(s)}(\tau,\tau')=
  \avr{\varphi_i(\tau)\varphi_j(\tau')}_s
\notag\\
&
=\Psi_{im}^{(\mu)}(\tau) D_{mn}^{(\mu\nu)}\Psi_{nj}^{(\nu)}(\tau')
\end{align}
where
$
D_{mn}^{(\mu\nu)}\equiv\avr{\varphi_m^{(\mu)}\varphi_n^{(\nu)}}_s$.
In our case, we can easily carry out the averaging over fluctuations 
by calculating  Gaussian integrals with 
the probability distribution defined in
Eqs.~\eqref{eq:prob-surf} and~\eqref{eq:en-surf}.
The result is
\begin{align}
  \label{eq:D-mn}
  \hat{D}\equiv
\begin{pmatrix}
\hat{D}^{(++)} & \hat{D}^{(+-)}\\
\hat{D}^{(-+)} & \hat{D}^{(--)}
\end{pmatrix}=
\frac{k_B T d}{ K_1 S}\,\hat{N}^{-1}.
\end{align}
In Appendix~\ref{sec:append-b} the Green function method is found to
yield the same result.

By using Eqs.~\eqref{eq:Psi-pm}
and~\eqref{eq:N} the surface part of the correlator
can now be derived from Eqs.~\eqref{eq:avr-surf}--\eqref{eq:D-mn}
in the final form: 
\begin{align}
  \label{eq:surf-corr-nlc}
&
  \hat{C}^{(s)}(\tau,\tau')=
\frac{k_B T d}{ 2 K_1 S}\,\hat{F}(\tau,\tau'),
\\
\label{eq:g-surf}
&
\hat{F}(\tau,\tau')=
\hat{\Psi}^{(s)}(\tau)\hat{M}_s^{-1}\hcnj{[\hat{\Psi}^{(s)}(\tau')]}
\notag\\
&
+
\hat{\Psi}^{(a)}(\tau)\hat{M}_a^{-1}\hcnj{[\hat{\Psi}^{(a)}(\tau')]}.
\end{align}
As is seen from Eq.~\eqref{eq:g-surf}, the surface part of the
correlator is a sum of two terms that represent 
the contributions coming from the symmetric and the antisymmetric fluctuation
harmonics at the substrates. 
The corresponding terms on the right hand side of
Eq.~\eqref{eq:en-s-a} provide expressions for
the free energies of these harmonics.

We can further emphasize the role of the matrices $\hat{M}_s$
and $\hat{M}_a$. For this purpose, we consider the
case in which $\tau=\tau'=\pm u$.
Since $\hat{C}^{(b)}(\pm u,\pm u)=0$, 
the covariance matrix of the director fluctuations at the
upper and lower substrates is determined by
the correlator~\eqref{eq:surf-corr-nlc}:
\begin{align}
  \label{eq:covar-matr}
&
 \hat{C}^{(s)}(\pm u,\pm u)=
 \hat{C}(\pm u,\pm u)
\notag\\
&
=
\begin{pmatrix}
\avr{\phi^2} & \avr{\theta\phi}\\
\avr{\theta\phi} & \avr{\theta^2}
\end{pmatrix}\Biggr\vert_{z=\pm d/2}. 
\end{align}
From Eq.~\eqref{eq:g-surf} we have
\begin{align}
  \label{eq:g-at-subs-p}
 \hat{F}(u,u)= \hat{M}_s^{-1} + \hat{M}_a^{-1},
\\
  \label{eq:g-at-subs-m}
 \hat{F}(-u,-u)= \hat{P} (\hat{M}_s^{-1} + \hat{M}_a^{-1}) \hat{P}. 
\end{align}
The fluctuations at the substrates are thus entirely 
described by the matrices $\hat{M}_s$
and $\hat{M}_a$. 
In Sec.~\ref{sec:direct-fluct-near}
we shall discuss symmetric and antisymmetric fluctuations
at the lower substrate of the cell
and present some numerical results for the elements of the matrix 
\begin{equation}
  \label{eq:f-alp}
 \hat{P} \hat{M}_{\alpha}^{-1} \hat{P}=
\begin{pmatrix}
F_{\phi\phi}^{(\alpha)} & F_{\theta\phi}^{(\alpha)}
\\
F_{\theta\phi}^{(\alpha)} & F_{\theta\theta}^{(\alpha)}
\end{pmatrix}.  
\end{equation}
From Eqs.~\eqref{eq:surf-corr-nlc} and~\eqref{eq:g-at-subs-m}
this matrix is proportional to
the contribution of the corresponding fluctuation mode to
the covariance matrix~\eqref{eq:covar-matr}.

\section{Stability of planar structure}
\label{sec:stab-pl-str}

It is now our task to study the stability of the planar structure.
In the preceding section we have derived all the analytical results required to
perform analysis of the stability
conditions~\eqref{eq:stab-crit} efficiently.
In our case violating these conditions will render 
the surface part of the correlator $\hat{C}^{(s)}$
divergent because the free energy~\eqref{eq:en-surf} 
is not positive definite and the corresponding Gaussian integrals
diverge. 

Equivalently, Eqs.~\eqref{eq:N} and~\eqref{eq:en-s-a} show that 
the planar structure is stable only if
the matrices $\hat{M}_s$ and $\hat{M}_a$ are both positive definite.
This yields the following two stability conditions:
\begin{equation}
  \label{eq:stab-det}
  \det\hat{M}_s > 0\, \text{ and }\,
\det\hat{M}_a > 0,
\end{equation}
that determine stability with respect the symmetric and the
antisymmetric fluctuation modes.

In order to proceed with our analysis we shall need to write down the
expressions for the determinants. From Eqs.~\eqref{eq:M_s}
and~\eqref{eq:M_a} we have
\begin{align}
  \label{eq:det-Ms}
&
  \det\hat{M}_s=
w_{\theta}w_{\phi}+\frac{(1-\rho) u}{\beta_s(u)}
(w_{\theta}\tanh^2 u + w_{\phi})
\notag\\
&
- u^2 q_{24}\Bigl(
q_{24}-2\,\frac{(1-\rho)\tanh u}{\beta_s(u)}
\Bigr),
\\
  \label{eq:det-Ma}
&
  \det\hat{M}_a=
w_{\theta}w_{\phi}+\frac{(1-\rho) u}{\beta_a(u)}
(w_{\phi}\tanh^2 u + w_{\theta})
\notag\\
&
- u^2 q_{24}\Bigl(
q_{24}-2\,\frac{(1-\rho)\tanh u}{\beta_a(u)}
\Bigr).
\end{align}
The determinants are written in the form of a sum of three terms, so
that only the last term can be negative 
(the functions $\beta_{s,a}(u)$ are defined in Eq.~\eqref{eq:bet-a-s}
and cannot be negative). 

This term is always negative at $q_{24}<0$ and 
we concentrate on the case of our primary concern in which
$q_{24}$ is positive. 
In this case  the term will be negative if the inequality
\begin{align}
  \label{eq:instb-cond}
  q_{24} > \gamma_{\alpha}(u)\equiv
2\,\frac{(1-\rho)\tanh u}{\beta_{\alpha}(u)}
\end{align}
is fulfilled. Given the value of $q_{24}$,
the values of the parameter $u$ that satisfy 
the instability condition~\eqref{eq:instb-cond}
form the instability interval for
the corresponding fluctuation harmonics.

\subsection{General method}
\label{sec:stab-anal}

The key point underlying our analysis 
is that the determinant $\det\hat{M}_{\alpha}$ 
will be negative provided the cell is sufficiently thin
and the parameter $u$ lies within the instability interval.
The reasoning is as follows.

Given the parameter $u$ ($=k_yd/2$), that meets the
condition~\eqref{eq:instb-cond}, 
the thickness of the cell $d$,
that enters Eqs.~\eqref{eq:det-Ms}--\eqref{eq:det-Ma}
through the parameters $w_{\theta}=d/(2l_{\theta})$ and
$w_{\phi}=d/(2l_{\phi})$, can be changed independently by varying
the fluctuation wavenumber $k_y$ so as to keep the parameter $u$
fixed. By this means the first two positive terms on the right hand sides
of Eqs.~\eqref{eq:det-Ms}--\eqref{eq:det-Ma} can be reduced to the
limit where the sign of the total sum is determined by
the last negative term.

From the above discussion it follows that
each value of $u$ from the instability interval 
defines the critical point where the determinant of the critical mode
vanishes. This point is characterized by the thickness of the cell
and the fluctuation wavelength, $\lambda=2\pi/k_y$.
The thickness can be computed as a function
of $u$ by finding the 
positive root of the quadratic equation $\det\hat{M}_{\alpha}=0$.
The wavelength then can be found from the relation 
$u=k_y d/2=\pi d/\lambda$ (see Eq.~\eqref{eq:notat-w}).

Geometrically, these calculations give a set of points  
$(d,\lambda)$ in the thickness--wavelength plane and
by this means the instability interval 
turned out to be mapped  
onto the curve $\Gamma_{\alpha}$ in the $d$--$\lambda$ plane.
This curve represents the boundary of the
instability region for the corresponding fluctuation harmonics.

It is now rather straightforward 
to carry out the described procedure and derive
the parametric equations for
the curve $\Gamma_{\alpha}$ in the form 
\begin{align}
  \label{eq:Gamma-alp}
&
  \Gamma_{\alpha}=
\begin{cases}
D=x_{\alpha}(u)=u\lambda_{\alpha}(u)/\pi,
\\
\Lambda = \lambda_{\alpha}(u),
\end{cases}
\end{align}
\begin{align}
  \label{eq:lmbd-s}
&
  \lambda_s(u)=\frac{\pi}{r_w\beta_s(u)}
\Bigl\{
-(1-\rho)(\tanh^2u+ r_w)
\notag\\
&
+
[(1-\rho)^2(\tanh^2u+ r_w)^2
\notag\\
&
+4r_w\beta_s(u) t_s(u)]^{1/2}
\Bigr\},
\end{align}
\begin{align}
  \label{eq:lmbd-a}
&
  \lambda_a(u)=\frac{\pi}{r_w\beta_a(u)}
\Bigl\{
-(1-\rho)(r_w\tanh^2u+ 1)
\notag\\
&
+
[(1-\rho)^2(r_w\tanh^2u+ 1)^2
\notag\\
&
+4r_w\beta_a(u) t_a(u)]^{1/2}
\Bigr\},
\end{align}
where
\begin{align}
\label{eq:t-alp}
&
t_{\alpha}(u)=q_{24}\beta_{\alpha}(u)
\bigl(
 q_{24}-\gamma_{\alpha}(u)
\bigr),
\\
  \label{eq:rw-DL}
&
D=\frac{d}{l_{\theta}},\quad
\Lambda=\frac{\lambda}{l_{\theta}},\quad
  r_w= \frac{W_{\phi}}{W_{\theta}},
\end{align}
$D$ is the size parameter; $\Lambda$ is the wavelength parameter;
and $r_w$ is the azimuthal anchoring parameter.

In Eqs.~\eqref{eq:Gamma-alp}--~\eqref{eq:t-alp} 
the thickness and the wavelength are 
both conveniently scaled by the zenithal anchoring extrapolation
length, $l_{\theta}$, 
and we shall use these formulas to perform numerical analysis
later on. But such analysis is not required for our subsequent
discussion.

Eqs.~\eqref{eq:Gamma-alp}--~\eqref{eq:t-alp} describe the boundary
of the instability region provided the instability interval is
not empty. So, we turn back to
the instability condition~\eqref{eq:instb-cond} to 
specify the regions of $q_{24}$ in which the instability may occur.
To this end we discuss the behavior of the functions
$\gamma_s(u)$ and $\gamma_a(u)$ that enter the right hand side of the
inequality~\eqref{eq:instb-cond}. The results, illustrated in
Fig.~\ref{fig:gamma}, are not difficult to obtain analytically.

Fig.~\ref{fig:gamma}(a) shows that, if  
the elastic constant $K_2$ is smaller than $K_1$ and
$r=K_2/K_1<1$, 
the function $\gamma_a$ increases from $\gamma_a(0)=2r$
asymptotically approaching the value $\gamma_a(\infty)=4r/(r+1)$ at large $u$,
while $\gamma_s$ is a decreasing function that varies from
$\gamma_s(0)=2$ to $\gamma_s(\infty)=4r/(r+1)$.
As is seen from Fig.~\ref{fig:gamma}(b), in the case where $r>1$
the functions $\gamma_a$ and $\gamma_s$ also
monotonically approach the common asymptotic value
$\gamma_a(\infty)=\gamma_s(\infty)=4r/(r+1)$. 
But now we have $\gamma_a(0)=2r>\gamma_s(0)=2$.

At the later stage we will see that there are two qualitatively different
regimes of instability depending on whether the instability interval
is bounded from above or not. 
So, we have the two characteristic values of $q_{24}$: 
\begin{align}
  \label{eq:qc-1}
&
q_c^{(1)}=\min(2,2 r),
\\
  \label{eq:qc-2}
&
q_c^{(2)}=2(1-\rho)=\frac{4 r}{r+1}.
\end{align}
Now we pass on to discuss the stability of the planar structure
in the intervals separated by these values.

\subsection{Stability diagrams}
\label{sec:stab-diagram}

In this section we study
the stability of the planar structure by
analyzing the behavior of the functions
which define the boundary curve~\eqref{eq:Gamma-alp} 
of the instability region.
Qualitatively, the analysis can be performed without resorting
to numerical computations and we shall use the numerical results 
only for illustrative purposes.

Fig.~\ref{fig:gamma} clearly indicates the interval
\begin{equation}
  \label{eq:Erick}
  0<q_{24}<q_c^{(1)}
\end{equation}
as the region where the inequalities~\eqref{eq:instb-cond}
cannot be satisfied and the structure is stable.
The stability conditions~\eqref{eq:Erick} 
are equivalent to the long known Ericksen inequalities~\eqref{eq:Erick-orig}
derived in Ref.~\cite{Eric:1966}. 
Fig.~\ref{fig:stab} shows
this stability region in the $r$--$q_{24}$ plane.

Next we consider the  interval
\begin{equation}
  \label{eq:part-instb}
  q_c^{(1)} <q_{24}< q_c^{(2)},
\end{equation}
where, as is demonstrated in Fig.~\ref{fig:gamma}(a), 
the instability takes place
at $u<u_{\smx}$ for the critical fluctuation mode
which symmetry depends on the parameter $r$. 
For relatively small elastic constants $K_2$ with
$r<1$, this mode is
antisymmetric (see Fig.~\ref{fig:gamma}(a)).
Referring to Fig.~\ref{fig:gamma}(b), it can be seen that
in the opposite case with $r>1$ the critical mode is symmetric.

Since the instability interval is bounded, 
the size parameter $D$ on the boundary
curve~\eqref{eq:Gamma-alp} is described by 
the function $x_{\alpha}(u)$ and reaches its maximum,
$D_c$, as $u$ varies from $0$ to $u_{\smx}$.
It follows that in this regime the planar structure
is unstable only in sufficiently thin films
and $d_c$ ($D_c=d_c/l_{\theta}$)
is the critical thickness.
On the curve $\Gamma_{\alpha}$ this critical point is also
characterized by the critical fluctuation wavelength
$\lambda_c$ ($\Lambda_c=\lambda_c/l_{\theta}$).

Since $q_{24}=\gamma_{\alpha}$ at $u=u_{\smx}$, 
Eqs.~\eqref{eq:det-Ms} and~\eqref{eq:det-Ma}
show that the $q_{24}$ dependent term
vanishes at $u=0$ and $u=u_{\smx}$.
When $W_{\theta}W_{\phi}\ne 0$,
the result is 
that $x_{\alpha}(0)=x_{\alpha}(u_{\smx})=0$ 
and the curve 
$\Gamma_{\alpha}$ forms the loop that encloses the region of
instability. Fig.~\ref{fig:dc-lmbc-1}(a) illustrates this result for
$r=1.5$ when the boundary curve $\Gamma_s$
is determined by the symmetric critical fluctuation mode. 

Now we consider the case of large $q_{24}$ with
\begin{equation}
  \label{eq:instab}
  q_{24}>q_c^{(2)}.
\end{equation}
In this case
the instability interval is no longer bounded from above.
From Fig.~\ref{fig:gamma} it is clear that
the instability condition~\eqref{eq:instb-cond}
does not impose any restrictions on $u$ for the critical mode,
whereas the instability interval for the non-critical mode is:
$u>u_{\smn}$.
Thus, the fluctuation modes both lead to the instability
at sufficiently large $u$.
Eqs.~\eqref{eq:lmbd-s}--\eqref{eq:t-alp}
provide the limiting value of 
the functions $\lambda_s(u)$ and $\lambda_a(u)$
at $u\to \infty$: 
\begin{align}
  \label{eq:lmb-asymp}
  &
  \Lambda_{\infty}=\frac{\pi}{r_w}
\Bigl\{
-(1-\rho)(r_w+ 1)
+
[(1-\rho)^2
\notag\\
&
\times
(r_w+ 1)^2
+4r_w q_{24}(q_{24}-q_c^{(2)})]^{1/2}
\Bigr\},
\end{align}
so that the size parameter, $D=u\lambda_{\alpha}/\pi$, 
increases indefinitely in this limit.
The result is that the planar structure is unstable
at any thickness of the cell and
the instability is caused by the short wavelength
fluctuations with $\lambda<\lambda_{\infty}$.
This result equally applies to the case of negative
$q_{24}$.

It should be emphasized that using the continuum
elastic theory we are limited to the case 
where the characteristic length of director distortions
is large as compared to a microscopic scale 
of intermolecular distances. 
This limitation can be taken into consideration by eliminating
the fluctuation harmonics with large wavenumbers,
$k_{x,y} > k_{\uv}=2\pi/\lambda_{\uv}$, 
from the Fourier series expansion~\eqref{eq:pl-Four},
so that the wavenumber $k_{\uv}$
gives the large momentum (ultraviolet) cut-off in the theory.
In this approach destabilizing effect of 
the critical short wavelength fluctuations
may only be indicated if $\lambda_{\infty}$ is larger than $\lambda_{\uv}$.
Our results are scaled by the zenithal anchoring extrapolation length,
$l_{\theta}$. Assuming that the anchoring is sufficiently weak and 
$l_{\theta}\gg\lambda_{\uv}$, we conclude that 
in this case the effect of the cut-off on
the accuracy of our results is negligible.

Figs.~\ref{fig:dc-lmbc-1}(b-c) illustrate transformations of the
stability diagram in the $D$--$\Lambda$ plane
as $q_{24}$ increases beyond $q_c^{(2)}$ at $r>1$.
The curve  of the critical mode $\Gamma_{s}$
forms the boundary of the instability region
similarly to the case of the interval~\eqref{eq:part-instb}.
The curve of the non-critical mode $\Gamma_{a}$
resides within this region
and is not shown in the figures.
Both curves rapidly approach the horizontal asymptote 
$\Lambda=\Lambda_{\infty}$.

We show in Fig.~\ref{fig:dc-lmbc-1}(b) that,
when $q_{24}$ is close to $q_c^{(2)}$,
the shape of 
the curve $\Gamma_s$ still somewhat resembles the loop developed in the
interval~\eqref{eq:part-instb}.
There are two points marked as $D_{\smn}$ and $D_{\smx}$ 
at which the curve bends and at which the function $x_s(u)$ reaches
its local minimum and maximum, correspondingly.
Further increase of $q_{24}$ reduces the distance between
$D_{\smn}$ and $D_{\smx}$ and the points finally disappear
after merging at sufficiently large $q_{24}$.
Fig.~\ref{fig:dc-lmbc-1}(c) presents the stability diagram
in this regime.

In general, Fig.~\ref{fig:stab} qualitatively summarizes the results
of this section for the
three different intervals of $q_{24}$ described in 
Eqs.~\eqref{eq:Erick}--\eqref{eq:instab}.
The interval~\eqref{eq:instab} was previously indicated as the
instability region in Ref.~\cite{Kis:mclc:1998}
and as the region where modulated structures can exist in
Ref.~\cite{Barb:pre4:2002}.

Our results for the interval~\eqref{eq:part-instb} show that
in this case the instability induced by the
$K_{24}$ term in cells of subcritical thickness
may also lead to the formation of periodically modulated 
structures. In addition, 
the critical wavelength $\lambda_c$ provides an estimate 
for the period of the emerging structure near the critical point. 
  
\subsection{Effects of azimuthal anchoring}
\label{sec:effects-azim-anch}

The critical thickness and the critical wavelength are meaningful
only when $q_{24}$ is within the interval~\eqref{eq:part-instb}.
By using Eqs.~\eqref{eq:Gamma-alp}--\eqref{eq:t-alp}
they both can be computed numerically. The numerical procedure
involves two steps: (a)~solving the equation $x'_{\alpha}(u_c)=0$
to find the maximum of $x_{\alpha}$ for the
critical mode; and (b)~evaluating $D_c=x_{\alpha}(u_c)$
and $\Lambda_c=\lambda_{\alpha}(u_c)$.
We also present the results for $\Lambda_{\infty}$
computed from Eq.~\eqref{eq:lmb-asymp}.

The critical thickness parameter, $D_c$,
the critical wavelength parameter, $\Lambda_c$, and
the parameter $\Lambda_{\infty}$ as functions of $q_{24}$
for different values of 
the azimuthal anchoring parameter, $r_w=W_{\phi}/W_{\theta}$,
are plotted in Figs.~\ref{fig:d-q24},~\ref{fig:l-q24}
and~\ref{fig:linf-q24}. 
As it can be expected, the curves demonstrate that the structure
becomes less stable as the azimuthal anchoring strength decreases.
So, in order to estimate the critical thickness from above,
it is instructive to discuss the limit of weak azimuthal anchoring,
$W_{\phi}\to 0$. This limiting case was considered in
Refs.~\cite{Barb:pre3:2002,Barb:pre4:2002} and we 
clarify some of the results by using our approach.
 
It is, however, should be stressed that
in this case the planar structure is marginally stable 
with respect to the long wavelength symmetric fluctuations with $k_y=0$
regardless of $K_{24}$. 
Mathematically, it follows because 
the determinant~\eqref{eq:det-Ms} vanishes at $u=0$ and $w_{\phi}=0$.
The reason is that
all uniform planar structures are energetically equivalent
in the absence of azimuthal anchoring.

Nevertheless, we may apply our method
to this case by taking the limit $w_{\phi}\to 0$
in Eqs.~\eqref{eq:Gamma-alp}--\eqref{eq:lmb-asymp}.
In this limit the expressions~\eqref{eq:lmbd-s},~\eqref{eq:lmbd-a}
and~\eqref{eq:lmb-asymp}
assume the simplified form
\begin{align}
  \label{eq:lmbs-rw0}
&
  \lambda_s(u)=\frac{2\pi\, t_s(u)}{(1-\rho)\tanh^2 u},
\\
  \label{eq:lmba-rw0}
&
  \lambda_a(u)=\frac{2\pi\, t_a(u)}{(1-\rho)},
\\
  \label{eq:l-inf-rw0}
&
  \Lambda_{\infty}=\lambda_s(\infty)
=\frac{2\pi q_{24}}{1-\rho} 
\bigl(q_{24}-q_c^{(2)}
\bigr).
\end{align}
From Eq.~\eqref{eq:lmba-rw0} it is seen that the behavior of the antisymmetric
harmonics at $u=0$ does not differ from the case in which 
$w_{\phi}\ne 0$ and $\lambda_a(0)=x_a(0)=0$.
But, for the symmetric mode, this is not the case. 
From Eq.~\eqref{eq:lmbs-rw0} we have
$\lambda_s\to \infty$ at $u\to 0$ and
\begin{align}
  \label{eq:dc-rw0}
&
D_c = x_s(0)= 2 q_{24}(q_{24}-2).
\end{align}
When $r>1$ and the critical mode is symmetric, Eq.~\eqref{eq:dc-rw0}
provides the exact expression for the critical thickness at
$W_{\phi}=0$. In Refs.~\cite{Barb:pre3:2002,Barb:pre4:2002}
the result~\eqref{eq:dc-rw0} has been derived as an approximation. 

The stability diagrams in the $D$--$\Lambda$ plane
at $r>1$ and $W_{\phi}=0$ are shown in Fig.~\ref{fig:dc-lmbc-0}.
In the opposite case of $r<1$, so long as $q_{24}<2$ 
the diagrams are quite similar to those
depicted in Fig.~\ref{fig:dc-lmbc-1}. Otherwise, 
at $q_{24}>2$, the non-critical symmetric mode
will change the stability diagram presented in
Fig.~\ref{fig:dc-lmbc-1}(c).
The result, however, is not too different from the diagram
shown in Fig.~\ref{fig:dc-lmbc-0}(c).

If the azimuthal anchoring strength is not identically zero, 
Eq.~\eqref{eq:dc-rw0} estimates the critical thickness from above.
In particular, substituting $q_c^{(2)}$ into Eq.~\eqref{eq:dc-rw0}
and taking the limit $r\to\infty$ we arrive at the conclusion that
the critical thickness cannot be larger than $16 l_{\theta}$.

Dependencies of the critical thickness, $D_c$, 
and the critical wavenumber, $K_c=2\pi/\Lambda_c$, 
on the ratio $r_w=W_{\phi}/W_{\theta}$ for various values of $r$
are plotted in Figs.~\ref{fig:d-rw} and~\ref{fig:k-rw}.
As is seen from Fig.~\ref{fig:d-rw}, the critical thickness declines
steeply in the immediate vicinity of the origin.
Typically, the critical thickness at $r_{w}=0$
appears to be halved at $r_w=0.1$, so that even for $r=20.0$
we need very small azimuthal anchoring $r_w\approx 0.03$ to have
$D_c\approx 8$ ($d_c\approx 8 l_{\theta}$).
Referring to Fig.~\ref{fig:k-rw}, the critical wavenumber, 
$K_c=2\pi/\Lambda_c$,
also start growing rapidly, 
but the effect is less pronounced at large values of
$r$. So, we have $K_c\approx 0.2$ ($k_c\approx 0.2/l_{\theta}$)
at $r_w\approx 0.03$ and $r=20.0$.

When the saddle-splay elastic constant meets a Cauchy
relation: $K_{24}=(K_1+K_2)/2$~\cite{Nehr:1971,Faet:jpfr:1995},
we find the two intervals for the twist-splay ratio where 
the regime of instability defined in Eq.~\eqref{eq:part-instb}
takes place:
$3<r<3+2\sqrt{2}\approx 5.82$ and
$3-2\sqrt{2}\approx 0.17<r<1/3$.
The critical thickness and the critical wavelength
as a function of the parameter $r$ 
varying within these intervals
are plotted in Fig.~\ref{fig:cauchy}.

For $r<1/3$, the critical fluctuation mode is antisymmetric
and Fig.~\ref{fig:cauchy}(c) shows that 
the critical wavelength remains finite at $r_w=0$.
Referring to Figs.~\ref{fig:cauchy}(a)-(b), 
the critical thickness at $r<1/3$ 
is an order of magnitude smaller
than in the case where $r>3$ and
the critical mode is symmetric.

According to Ref.~\cite{Barb:pre3:2002},
the latter presents the case in which the ratio $r$ 
grows anomalously large due to an increase of 
the twist constant $K_2$
in the vicinity of the nematic-smectic-$A$ transition. 
In this case, the estimate of 
the critical thickness at
$W_{\phi}=0.0$ and  
$q_{24}=q_c^{(2)}\approx 3.41$
($r\approx 5.82$)
provides the upper bound for $d_c$ to be about
$9.7\, l_{\theta}$.
As is shown in Fig.~\ref{fig:cauchy}(b),
this estimate can be significantly reduced in the presence
of azimuthal anchoring.

\subsection{Director fluctuations at substrates near the critical point}
\label{sec:direct-fluct-near}

In this section we apply the results of
Sec.~\ref{sec:direct-fluct-subs}
to study the correlation functions of director 
fluctuations at the plates bounding the cell.
Specifically, we shall use
Eqs.~\eqref{eq:covar-matr}--\eqref{eq:f-alp} that express the
correlator in terms the inverse of the matrices
$\hat{M}_s$ and $\hat{M}_a$ given in Eqs.~\eqref{eq:M_s}
and~\eqref{eq:M_a}. 

Obviously, the results obtained in the Gaussian
approximation are inapplicable in the region where the structure is
unstable. So, we shall restrict our considerations to the stability
region and study what happen when 
either the thickness or $q_{24}$ varies so as to
get closer to the boundary of the instability region.

We start with the discussion of the long wavelength limit,
$u\to 0$, for the matrices $\hat{M}_{\alpha}^{-1}$.
The result of calculations is that these matrices are diagonal
at $u=0$:
$\hat{M}_s^{-1}=\diag(w_{\phi}^{-1},(w_{\theta}+1)^{-1})$
and
$\hat{M}_a^{-1}=\diag((w_{\phi}+r)^{-1},w_{\theta}^{-1})$.
It is seen that in this limit
the correlator diverges 
for vanishing the anchoring strengths, $W_{\phi}=0$ or $W_{\theta}=0$.
This is a consequence of the marginal instability discussed in
the preceding section.

As far as the large wavenumber (short wavelength) limit
is concerned, it can be shown that in the stability region 
the matrices (and the correlator) both decay to zero
at $u\to\infty$.
Interestingly, this is not the case 
at the boundary of the instability interval
when $q_{24}=0$ or $q_{24}=q_c^{(2)}$.
In this case we have the non-zero limits  
$\hat{M}_{\alpha}^{-1}(\infty)=(w_{\theta}+w_{\phi})^{-1} 
\begin{pmatrix}
1 & \pm 1\\
\pm 1 & 1
\end{pmatrix}
$
for $q_{24}=0$ and $q_{24}=q_c^{(2)}$, respectively.
This anomaly is a precursor of the instability
caused by short wavelength fluctuations.
As a result, using the widespread approximation with
$K_{24}=0$ does not give the correlation functions
that properly behave at large wavenumbers. 

We demonstrate in Figs.~\ref{fig:corr-1} and~\ref{fig:corr-02},
which show the spectra of the critical and the non-critical
fluctuation modes computed from Eq.~\eqref{eq:f-alp} at $r=1.5$, 
that the critical increase of  the symmetric fluctuation mode
becomes sharply peaked at the critical wavenumber as
the thickness of the cell approaches its critical value.
In addition, Fig.~\ref{fig:corr-1} shows that
the symmetric and the antisymmetric modes are 
dominated by the in-plane and by the out-of-plane fluctuations,
respectively.

\section{Discussion and Conclusions}
\label{sec:disc-concl}

In this paper stability of the uniform director distribution in a
planar NLC cell has been studied in the presence of the saddle-splay
term. We have shown that the physical mechanism behind instabilities
induced by the $K_{24}$ term is governed solely by the director
fluctuations at the substrates that
can generally be characterized by
the free energy and the probability distribution.
We have developed the approach to study such instabilities
by using the correlation function of 
the fluctuation field induced by the fluctuations at the confining wall
(the surface part of the correlator) for derivation of the stability
conditions.

This approach in combination with the mirror symmetry considerations has
been applied to the case of NLC planar cell. It is found that there
are two types of fluctuation modes, that we have called symmetric and
antisymmetric modes, depending on the parity under the mirror symmetry
transformation involving reflection in the middle plane of the cell.

We have devised the analytical method to analyze the stability
conditions for the fluctuation modes of different symmetry.
In this method the thickness of the cell and the fluctuation
wavelength form the plane and the boundary of the instability region
in this plane is described as a curve defined in the parametric
form. It appears that no numerical calculations is required 
as far as qualitative analysis of stability diagrams is concerned.
The analysis revealed the two different regimes for instabilities
caused by the saddle-splay term depending on the value of $K_{24}$.

When $K_{24}$ falls within the range between 
$ 2\min(K_1,K_2)$ and $4K_1K_2/(K_1+K_2)$, the planar
orientational structure loses its stability only in sufficiently
thin cells and the critical thickness, $d_c$, together with the
wavelength of the critical fluctuation mode, $\lambda_c$, determine
the critical point. In this case, as is shown in
Fig.~\ref{fig:corr-02}, the spectrum of critical fluctuations at the
substrates grows sharply peaked at the critical wavelength when
approaching the critical point.  The period of modulated structure
emerging at the critical point is thus determined by the critical
wavelength.

This $K_{24}$ induced instability takes place at any elastic anisotropy
parameter $r=K_2/K_1$. The sole exception is the case of elastic isotropy
in which $K_1=K_2$. 
In contrast with 
the periodic splay-twist Fr\'{e}edericsz 
transition~\cite{Lonb:1985,Zim:1986,Miral:1986},
where spatially modulated pattern comes into play only at sufficiently
small twist-splay ratio with $K_2/K_1$, 
the surface elasticity driven instability
cannot be hindered by large elastic anisotropy, but rather,
as is shown in Fig.~\ref{fig:d-rw}, 
the greater elastic anisotropy the larger
the critical thickness can be.
The symmetry of the critical fluctuation mode, however,
depends on the parameter $r$: the mode is antisymmetric at $r<1$
and is symmetric in the opposite case of $r>1$.
Fig.~\ref{fig:corr-1} shows that the in-plane and out-of-plane
fluctuations prevail depending on the symmetry of the fluctuation mode.

The azimuthal anchoring turned out to have a profound effect on both
the critical thickness and the critical wavelength. 
The absence of the azimuthal anchoring
presents the limiting case where the planar structure
is marginally unstable with respect to the long wavelength
fluctuations with $k_y=0$ regardless of the $K_{24}$ term. 
As a consequence, the critical wavelength $\lambda_c$
increases indefinitely in the limit of weak azimuthal anchoring,
$\lambda_c\to \infty$ at 
$W_{\phi}\to 0$, while the limiting value of the critical thickness
can be computed exactly (see Eq.~\eqref{eq:dc-rw0})
giving the upper bound for the critical thickness.

In Sec.~\ref{sec:effects-azim-anch}, 
the absolute upper bound for
the critical thickness was found to be $16\, l_{\theta}$.
Fig.~\ref{fig:d-rw} shows that the critical thickness
could have been significantly reduced in the presence of
relatively small amount of the azimuthal anchoring.

Using the Cauchy
relation: $K_{24}=(K_1+K_2)/2$~\cite{Nehr:1971,Faet:jpfr:1995},
we have found that the instability may
occur at both sufficiently low and high twist-splay ratios 
with $r<1/3$ and $r>3$, respectively
(see Fig.~\ref{fig:cauchy} and the related discussion at the end
of Sec.~\ref{sec:effects-azim-anch}).

The corresponding condition for
the periodic Fr\'{e}edericsz transition requires the ratio
$r$ to be below $r_c\approx 0.303$~\cite{Zim:1986,Miral:1986}
which places a slightly more stringent constraint on the value of $r$ 
than the above inequality: $r<1/3$.
But, as is seen from Fig.~\ref{fig:cauchy}(a),
for small $r$, the critical thickness is 
an order of magnitude smaller
than in the case of large $r$ presented in Fig.~\ref{fig:cauchy}(b),
where $d_c$ can be of order of several microns
provided the extrapolation length $l_{\theta}$ varies in the range
$0.1-1\,\mu$m.

For typical nematics, the twist-splay ratio does not exceed unity, but
close to the nematic-smectic-$A$ transition
the parameter $r$ becomes anomalously large~\cite{Gennes:bk:1993}
leading to the $K_{24}$ induced instability of the ground state
at $r>3$. This may result in the appearance of modulated orientational
structures as suggested in Ref.~\cite{Barb:pre3:2002}.

When $ K_{24}>4K_1K_2/(K_1+K_2)$ or $K_{24}<0$, the short wavelength
fluctuations with $\lambda<\lambda_{\infty}$ will render the planar
structure unstable at any thickness of the cell.  With the Cauchy
relation such instability will take place when the twist-splay ratio
is either less than $0.17$ or greater than $5.82$.  In contrast with
the above discussed regime, this instability though does not impose
any restrictions on the film thickness, in general, cannot be
unambiguously related to the periodic pattern formation.  This case
requires a more detailed additional study of orientational structures
in the instability region where the Gaussian approximation is
inapplicable.

Our concluding remark concerns the general method for separating out
the contribution of director fluctuations at confining walls to static
correlation functions.  We have demonstrated that this method can be
used as a useful tool for studying orientational instabilities in
confined liquid crystals. For this purpose, we have restricted
ourselves to the case of uniaxial director fluctuations with uniformly
distributed degree of ordering. 
But a complete
treatment of fluctuations at confining surfaces is required in studies of
such phenomena as
electrohydrodynamical pattern formation~\cite{Buka:bk:1996},
instabilities under shear flow~\cite{Gennes:bk:1993,TJS:prs:2003}, 
wetting~\cite{Zih:pre:1998} and backflow~\cite{Gennes:bk:1993}.
These more general considerations involving
fluctuations of spatially varying order parameter
tensor coupled to the translational degrees of freedom
is well beyond the scope of this
paper and we will extend on this subject elsewhere.

\begin{acknowledgments}
The author thanks V.M.~Pergamenshchik for useful discussions.
\end{acknowledgments}

\appendix

\section{Free energy of fluctuations in nematic cell}
\label{sec:append}

In this appendix we briefly comment on the derivation of 
the expression~\eqref{eq:F2m} for the free energy.
We start with the second order 
expression for the free energy~\eqref{eq:happr-2}: 
\begin{align}
\label{eq:A-fb-2}
&
  F_b^{(2)}=\frac{1}{2}\, \int_V\Bigl[
K_1\, (\prt{y}\phi+\prt{z}\theta)^2+
K_2\,(\prt{y}\theta-\prt{z}\phi)^2
\notag\\
&
+
K_3\,\bigl[(\prt{x}\phi)^2+(\prt{x}\theta)^2
\bigr]
\Bigr]\dd v,
\end{align}
\begin{align}
\label{eq:A-fs-2}
&
F_s^{(2)}=\frac{1}{2}\, \int_S\Bigl[
W_{\phi}\phi^2+W_{\theta}\theta^2
\notag\\
&
-
K_{24}\,\bigl[\theta\,\prt{y}\phi
-
\phi\,\prt{y}\theta
\bigr]
\Bigr]\dd s,
\end{align} 
where we retain the director derivatives with respect to $x$ 
for discussing its role later on.

Substituting the expansion~\eqref{eq:pl-Four}
into Eqs.~\eqref{eq:A-fb-2} and~\eqref{eq:A-fs-2}
will provide the free energy of the fluctuation harmonics
in the following matrix form
\begin{align}
  \label{eq:A-fm}
&
  F_{\vc{m}}^{(2)}/S=\Bigl(1-\frac{\delta_{0\vc{m}}}{2}\Bigr)
\int_{-d/2}^{d/2}\dd z
\Bigl[
\hcnj{\prt{z}\bs{\psi}}_{\vc{m}}\hat{A}_1\,\prt{z}\bs{\psi}_{\vc{m}}
\notag\\
&
+\hcnj{\bs{\psi}}_{\vc{}m}\hat{B}_1\prt{z}\bs{\psi}_{\vc{m}}
+
\hcnj{\prt{z}\bs{\psi}}_{\vc{m}}\hcnj{\hat{B}_1}\bs{\psi}_{\vc{m}}
+ \hcnj{\bs{\psi}}_{\vc{m}}\hat{C}_1\bs{\psi}_{\vc{m}}
\Bigr]
\notag\\
&
+\sum_{\mu=\pm 1}
\hcnj{\bs{\psi}}_{\vc{m}}\hat{D}_{\mu}\bs{\psi}_{\vc{m}}
\Bigr\vert_{z=\mu d/2}.
\end{align}
Using the transformation~\eqref{eq:notat-1} will render all matrices
in Eq.~\eqref{eq:A-fm} real-valued. These are given by
\begin{align}
  \label{eq:A-A1}
&
  \hat{A}_1=\begin{pmatrix} K_2 & 0\\ 0 & K_1\end{pmatrix},
\\
  \label{eq:A-B1}
&
\hat{B}_1=-k_y\begin{pmatrix} 0 & K_1\\ K_2 & 0\end{pmatrix},
\end{align}
\begin{align}
  \label{eq:A-C1}
&
\hat{C}_1=\begin{pmatrix} K_1 k_y^2 & 0\\ 0 & K_2k_y^2\end{pmatrix}
+K_3 k_x^2\hat{I},
\end{align}
\begin{align}
\label{eq:A-Dmu}
&
\hat{D}_{\mu}=\mu k_y
\begin{pmatrix} 0 & K_{24}\\ K_{24} & 0\end{pmatrix}
\notag\\
&
+
\begin{pmatrix} W_{\phi}^{(\mu)} & 0\\ 0 & W_{\theta}^{(\mu)}
\end{pmatrix},
\end{align}
where $W_{\phi}^{(\mu)}$ and $W_{\theta}^{(\mu)}$
are the azimuthal and the zenithal anchoring strengths at the
substrate $z=\mu d/2$. 

For a symmetric cell with identical substrates,
we have $W_{\phi}^{(\pm)}=W_{\phi}$ and
$W_{\theta}^{(\pm)}=W_{\theta}$.
This is the case we deal with in this paper.
But the generalized expression~\eqref{eq:A-Dmu}
can be used to discuss the effects of anchoring energy asymmetry.

Another generalization is that the derivatives with respect to $x$
have not been neglected from the very beginning and we have
the term proportional to $k_x^2$ in the expression for
the matrix~\eqref{eq:A-C1}. 

But the stabilizing magnetic field $\vc{H}=H\vc{e}_x$
likewise will change the matrix~\eqref{eq:A-C1}:
$\hat{C}_1\to\hat{C}_1+ \chi_a H^2\hat{I}$, 
where $\chi_a$ is the anisotropic part of the magnetic susceptibility.
It follows that the effects of non-vanishing wavenumbers $k_x$
and of the stabilizing magnetic field are equivalent, so that,
at $k_x\ne 0$, the structure is more stable
than in the case with $k_x=0$. Thus, the fluctuation harmonics with
$k_x\ne 0$ do not affect the results of stability analysis
and can be safely eliminated from the consideration.

Assuming that $k_x=0$ and the cell is symmetric, 
Eqs.~\eqref{eq:A-fm}--\eqref{eq:A-Dmu}
can now be rewritten in terms of 
the dimensionless parameters defined in Eqs.~\eqref{eq:notat-rq}
and~\eqref{eq:notat-w} to yield
the expressions~\eqref{eq:FFm}--\eqref{eq:Q-mu-def}.

\section{Green function method for nematic cell}
\label{sec:append-b}

In this appendix we deduce an expression for the
correlator of director fluctuations in the NLC cell 
by using the Green function formalism  described in
Sec.~\ref{subsec:corr-green-funct}.
We shall find that this method can also be employed to
rederive the result for the surface part of the correlator
given in Eqs.~\eqref{eq:avr-surf} and~\eqref{eq:D-mn}.
In addition, this appendix provides a number of technical details
omitted in the bulk of the paper.

We start with the expression for the free energy
functional~\eqref{eq:Smb}
taken, similarly to Eq.~\eqref{eq:Fb-repr}, in the following form
\begin{align}
  \label{eq:Smb-b}
&
  S_{m}^{(b)}[\bs{\psi}]=-u\,\int_{-u}^{u}
\hcnj{\bs{\psi}}\hat{L}\bs{\psi}\, \dd\tau
\notag\\
&
+u\sum_{\mu=\pm 1}
\hcnj{\bs{\psi}}\hat{Q}_{\mu}^{(b)}\bs{\psi}\Bigr\vert_{\tau=\mu u}\,,
\end{align}
\begin{align}
\label{eq:Q-mu-b}
&
\hat{Q}_{\mu}^{(b)}=\mu\Bigl(
\hat{A}\prt{\tau}-\frac{1}{2}\,\hat{B}
\Bigr).
\end{align}
Substituting the fluctuation field~\eqref{eq:sol-bc} 
expressed in terms of the matrices $\hat{\Psi}^{(\pm)}$
that satisfy the Euler-Lagrange equations~\eqref{eq:EU-Smb}
and meet the boundary conditions~\eqref{eq:Psi-pm-bc}
into Eqs.~\eqref{eq:Smb-b} and~\eqref{eq:Sms}
gives the matrix $\hat{N}$ that enters Eq.~\eqref{eq:en-surf}
\begin{align}
  \label{eq:N-tot-b}
&
  \hat{N}=
\begin{pmatrix}
\hat{N}_{++}& \hat{N}_{+-}\\
\hat{N}_{-+}& \hat{N}_{--}
\end{pmatrix},
\end{align}
where
\begin{align}
  \label{eq:N-mn-b}
&
\hat{N}_{\mu\nu}
=\hat{Q}_{\mu}\hat{\Psi}^{(\nu)}\Bigr\vert_{\tau=\mu u}=
u\hat{N}_{\mu\nu}^{(b)}
\notag\\
&
+\delta_{\mu\nu}\hat{Q}_{\mu}^{(s)},\:
\hat{Q}_{\mu}= u \hat{Q}_{\mu}^{(b)}+
\hat{Q}_{\mu}^{(s)},
\end{align}
\begin{align}
  \label{eq:N-b-b}
&
\hat{N}_{\mu\nu}^{(b)}
=\hat{Q}_{\mu}^{(b)}\hat{\Psi}^{(\nu)}\Bigr\vert_{\tau=\mu u}.
\end{align}

We define the Green function as the solution of the following boundary
value problem:
\begin{subequations}
\label{eq:Gr-prob-b}
\begin{align}
  \label{eq:Gr-eq-b}
&
  \hat{L}\,\hat{G}(\tau,\tau')=-\delta(\tau-\tau')\hat{I},
\\
  \label{eq:Gr-bc-b}
&
\hat{Q}_{\pm}\hat{G}\Bigr\vert_{\tau=\pm u}=0.
\end{align}
\end{subequations}
Since the operator $\hat{K}$ from Eq.~\eqref{eq:Green-eq}
is proportional to the operator $-\hat{L}$:
$\hat{K}=-uK_1 S/d\,\hat{L}$, we need to modify 
the relation~\eqref{eq:Green-corr-rel} linking
the Green function and the correlator as follows 
\begin{align}
  \label{eq:Cor-Gr-b}
&
  \hat{C}(\tau,\tau')=
\frac{k_B T d}{u K_1 S}\,\hat{G}(\tau,\tau').
\end{align}
Next we seek the Green function $\hat{G}$
in the form of a sum
\begin{align}
  \label{eq:Gr-decomp-b}
&
\hat{G}(\tau,\tau')=
\hat{G}^{(b)}(\tau,\tau')+
\hat{G}^{(s)}(\tau,\tau'),
\end{align}
where $\hat{G}^{(b)}$ is the Green function
of the Dirichlet problem: 
\begin{subequations}
\label{eq:Grb-prob-b}
\begin{align}
  \label{eq:Grb-eq-b}
&
  \hat{L}\hat{G}^{(b)}(\tau,\tau')=-\delta(\tau-\tau')\hat{I},
\\
  \label{eq:Grb-bc-b}
&
\hat{G}^{(b)}(\pm u,\tau')=0.
\end{align}
\end{subequations}
The boundary conditions~\eqref{eq:Grb-bc-b} of the Dirichlet problem
present the limiting case of strong anchoring and the Green function
$\hat{G}^{(b)}$ is proportional to the bulk part of the
correlator~\eqref{eq:Cor-Gr-b}.

Since the sum~\eqref{eq:Gr-decomp-b} is the solution of the 
problem~\eqref{eq:Gr-prob-b},
the equations and the boundary conditions for $\hat{G}^{(s)}$
are given by
\begin{subequations}
\label{eq:Grs-prob-b}
\begin{align}
  \label{eq:Grs-eq-b}
&
\hat{L}\hat{G}^{(s)}=0,
\\
  \label{eq:Grs-bc-b}
&
  \hat{Q}_{\mu}\hat{G}^{(s)}\Bigr\vert_{\tau=\mu u}=
-u \hat{Q}_{\mu}^{(b)}
\hat{G}^{(b)}\Bigr\vert_{\tau=\mu u}.
\end{align}
\end{subequations}
Clearly, the surface part of the Green function represented by $\hat{G}^{(s)}$
accounts for the difference between the boundary
conditions~\eqref{eq:Gr-bc-b} and the strong anchoring
conditions~\eqref{eq:Grb-bc-b}. 

We have already pointed out the difference
between the free energy~\eqref{eq:Fb-repr} and the
functional~\eqref{eq:Smb-b}
above the modified relation~\eqref{eq:Cor-Gr-b}.
Likewise, this difference will slightly change 
the Green formula~\eqref{eq:Green-forml} 
\begin{align}
  \label{eq:Green-form-b}
&
  \int_{-u}^{u}\Bigl[
\hcnj{\bs{\psi}}\hat{L}\bs{\varphi}-
\hcnj{\bs{\varphi}}\hat{L}\bs{\psi} 
\Bigr]
\dd\tau
\notag\\
&
=\sum_{\mu=\pm 1}
\Bigl[
\hcnj{\bs{\psi}}\hat{Q}_{\mu}^{(b)}\bs{\varphi}
-\hcnj{\bs{\varphi}}\hat{Q}_{\mu}^{(b)}\bs{\psi}
\Bigr]_{\tau=\mu u}\,.
\end{align}
There are two important relations that almost immediately follow from
the Green formula~\eqref{eq:Green-form-b}:
\begin{align}
  \label{eq:Grb-rel-b}
 -\hat{Q}_{\mu}^{(b)}
\hat{G}^{(b)}(\tau,\tau')\Bigr\vert_{\tau=\mu u}=
\hcnj{[\hat{\Psi}^{(\mu)}(\tau')]} 
\end{align}
and
\begin{align}
  \label{eq:Nb-sym-b}
  \hcnj{[\hat{N}_{\mu\nu}^{(b)}]}
=\hat{N}_{\nu\mu}^{(b)}.
\end{align}
Eq.~\eqref{eq:Grb-rel-b} can be derived from
Eq.~\eqref{eq:Green-form-b}
by setting $\bs{\psi}=\hat{G}^{(b)}(\tau,\tau')$ and
$\bs{\varphi}=\hat{\Psi}^{(\mu)}(\tau)\bs{\varphi}^{(\mu)}$,
whereas for the relation~\eqref{eq:Nb-sym-b} 
the substitution is:
$\bs{\psi}=\hat{\Psi}^{(\nu)}(\tau)\bs{\psi}^{(\nu)}$ and
$\bs{\varphi}=\hat{\Psi}^{(\mu)}(\tau)\bs{\varphi}^{(\mu)}$.

Combining Eqs.~\eqref{eq:N-tot-b}--\eqref{eq:N-b-b}
and the relation~\eqref{eq:Nb-sym-b} shows that the matrix
$\hat{N}$ is symmetric
\begin{align}
  \label{eq:N-sym-b}
  \hcnj{\hat{N}}
=\hat{N}.
\end{align}
So, the matrices that enter the right hand sides of
Eqs.~\eqref{eq:en-surf} and~\eqref{eq:D-mn} are identical.
In addition, the matrices $\hat{M}_{\alpha}$ defined in
Eq.~\eqref{eq:gen-M-sa} are also symmetric. The latter, however,
can be seen from the explicit formulas~\eqref{eq:M_alpha}.

Now we write the surface part of the Green function as the general
solution of Eq.~\eqref{eq:Grs-eq-b}
\begin{align}
  \label{eq:Grs-repr-b}
  \hat{G}^{(s)}(\tau,\tau')=
\sum_{\nu=\pm 1}
\hat{\Psi}^{(\nu)}(\tau)
\hat{\Theta}^{(\nu)}(\tau')
\end{align}
and compute the matrices $\hat{\Theta}^{(\nu)}(\tau')$ from the
boundary conditions~\eqref{eq:Grs-bc-b}.
The right hand side of Eq.~\eqref{eq:Grs-bc-b} can be simplified by
means of the relation~\eqref{eq:Grb-rel-b}.
Then substituting Eq.~\eqref{eq:Grs-repr-b} into
Eq.~\eqref{eq:Grs-bc-b}
and using Eq.~\eqref{eq:N-mn-b} yield the equations for
$\hat{\Theta}^{(\nu)}(\tau')$ in the final form
\begin{align}
  \label{eq:Grs-bc2-b}
\sum_{\nu=\pm 1}
\hat{N}_{\mu\nu}  \hat{\Theta}^{(\nu)}(\tau')= u
\hcnj{[\hat{\Psi}^{(\mu)}(\tau')]}. 
\end{align}
It can now be easily verified that 
Eqs.~\eqref{eq:Cor-Gr-b},~\eqref{eq:Gr-decomp-b},~\eqref{eq:Grs-repr-b} 
and~\eqref{eq:Grs-bc2-b} provide the expression for
the surface part of the correlator
given in Eqs.~\eqref{eq:avr-surf} and~\eqref{eq:D-mn}.

The relation~\eqref{eq:Grb-rel-b} can also be applied to derive
the bulk part of the Green function by means of the so-called
Wronski construction~\cite{Klein:bk:1999}. 
The result is
\begin{align}
  \label{eq:Grb-expr-b}
&
 \hat{G}^{(b)} =
\begin{cases}
\hat{\Psi}^{(+)}(\tau)\hat{R}
\hcnj{[\hat{\Psi}^{(-)}(\tau')]},
& \tau\le\tau'\\
\hat{\Psi}^{(-)}(\tau)\hcnj{\hat{R}}
\hcnj{[\hat{\Psi}^{(+)}(\tau')]},
& \tau\ge\tau' 
\end{cases},
\notag\\
&
\hat{R}=-[\hat{N}_{-+}]^{-1}. 
\end{align}

 Our final remark concerns the symmetry relation
 \begin{align}
   \label{eq:Psi-sym-b}
   \hat{\Psi}^{(-\mu)}(\tau)=
 \hat{P}\hat{\Psi}^{(\mu)}(-\tau)\hat{P}
 \end{align}
 that follows from Eqs.~\eqref{eq:Psi-p} and~\eqref{eq:Psi-m}
 and in combination with 
 Eqs.~\eqref{eq:N-mn-b} and~\eqref{eq:N-b-b} gives the following result
 \begin{subequations}
 \begin{align}
   \label{eq:Nb-sym2-b}
 &
   \hat{N}_{-\mu,\,-\nu}^{(b)}
 =\hat{P}\hat{N}_{\mu\nu}^{(b)}\hat{P},
 \\
   \label{eq:N-sym2-b}
 &
   \hat{N}_{-\mu,\,-\nu}
 =\hat{P}\hat{N}_{\mu\nu}\hat{P}.
 \end{align}
 \end{subequations}
 This result simplifies algebraic verification of the relation~\eqref{eq:N}. 


\end{document}